\RequirePackage{fix-cm}
\documentclass{svjour3}                     
\smartqed  
\usepackage{graphicx}
\usepackage{tablefootnote}
\usepackage{amsmath,amssymb,amsfonts}
\usepackage{algorithmic}
\usepackage{textcomp}
\usepackage{float}
\usepackage{subfig}
\usepackage{multirow}
\usepackage[table,xcdraw]{xcolor}
\usepackage{multirow}
\usepackage{adjustbox}
\usepackage{hhline}
\usepackage{pifont}
\usepackage{natbib}

\newcommand{\cmark}{\text{\ding{51}}}
\newcommand{\xmark}{\text{\ding{55}}}
\DeclareMathOperator*{\argmax}{arg\,max}

\journalname{Multimedia Tools and Applications}

\begin{document}

\title{Comprehensive Empirical Evaluation of Deep Learning Approaches for Session-based Recommendation in E-Commerce}


\titlerunning{Deep Learning Approaches for Session-based Recommendation in E-Commerce}

\author{Mohamed Maher \and Perseverance Munga Ngoy \and Aleksandrs Rebriks \and Cagri Ozcinar \and Josue Cuevas \and Rajasekhar Sanagavarapu \and Gholamreza Anbarjafari}

\authorrunning{Maher \textit{et al.}} 

\institute{Mohamed Maher \at
              University of Tartu, Tartu, Estonia \\
              \email{mohamed.abdelrahman@ut.ee}
           \and
           Perseverance Munga Ngoy, Aleksandrs Rebriks, Cagri Ozcinar and Gholamreza Anbarjafari \at
              iCV Lab, University of Tartu, Tartu, Estonia
           \and
           Gholamreza Anbarjafari \at
              Faculty of Engineering, Hasan Kalyoncu University, Gaziantep, Turkey
           \and
           Josue Cuevas, Rajasekhar Sanagavarapu \at
              Rakuten Inc., Big Data Department, Machine Learning Group, Tokyo, Japan
}

\date{Received: date / Accepted: date}

\maketitle

\begin{abstract}
Boosting sales of e-commerce services is guaranteed once users find more matching items to their interests in a short time. Consequently, recommendation systems have become a crucial part of any successful e-commerce services. Although various recommendation techniques could be used in e-commerce, a considerable amount of attention has been drawn to session-based recommendation systems during the recent few years. This growing interest is due to the security concerns in collecting personalized user behavior data, especially after the recent general data protection regulations \footnote{https://gdpr-info.eu/}. In this work, we present a comprehensive evaluation of the state-of-the-art deep learning approaches used in the session-based recommendation. In session-based recommendation, a recommendation system counts on the sequence of events made by a user within the same session to predict and endorse other items that are more likely to correlate with his/her preferences. Our extensive experiments investigate baseline techniques (\textit{e.g.,} nearest neighbors and pattern mining algorithms) and deep learning approaches (\textit{e.g.,} recurrent neural networks, graph neural networks, and attention-based networks). Our evaluations show that advanced neural-based models and session-based nearest neighbor algorithms outperform the baseline techniques in most of the scenarios. However, we found that these models suffer more in case of long sessions when there exists drift in user interests, and when there is no enough data to model different items correctly during training. Our study suggests that using hybrid models of different approaches combined with baseline algorithms could lead to substantial results in session-based recommendations based on dataset characteristics. We also discuss the drawbacks of current session-based recommendation algorithms and further open research directions in this field.
\end{abstract}




\section{Introduction}
\label{sec:introduction}
\label{intro}
Most of the e-commerce services use recommendation systems for helping their customers to find the items of interest based on their navigation behavior through these services. Recommendation systems are considered as a category of information filtering systems that aims to predict the users' preferences based on their behavior. They have become a crucial part of any successful business that helps satisfy user needs and boost the business sales volume \citep*{lee2014impact}. Recommendation systems have been used in a wide range of domains including images \citep*{liu2017related}, music \citep*{van2013deep}, videos \citep*{deldjoo2016content}, and even news \citep*{wang2018dkn} recommendations.

\label{literature}
Various types of recommendation systems have been proposed in literature, categorized as time-aware and session-based recommendation systems. The former can adapt to the temporal dynamics and user preferences drift over time \citep*{koren2009collaborative,song2016multi} and recommendation systems based on social information datasets \citep*{DBLP:journals/corr/abs-1902-07243,deng2016deep}. The latter relies on the user navigation behavior and sequence of actions and mouse clicks on different items solely to recommend the items that match the user's interest.

\begin{figure}[!ht]
    \centering
    \includegraphics[width=0.8\textwidth]{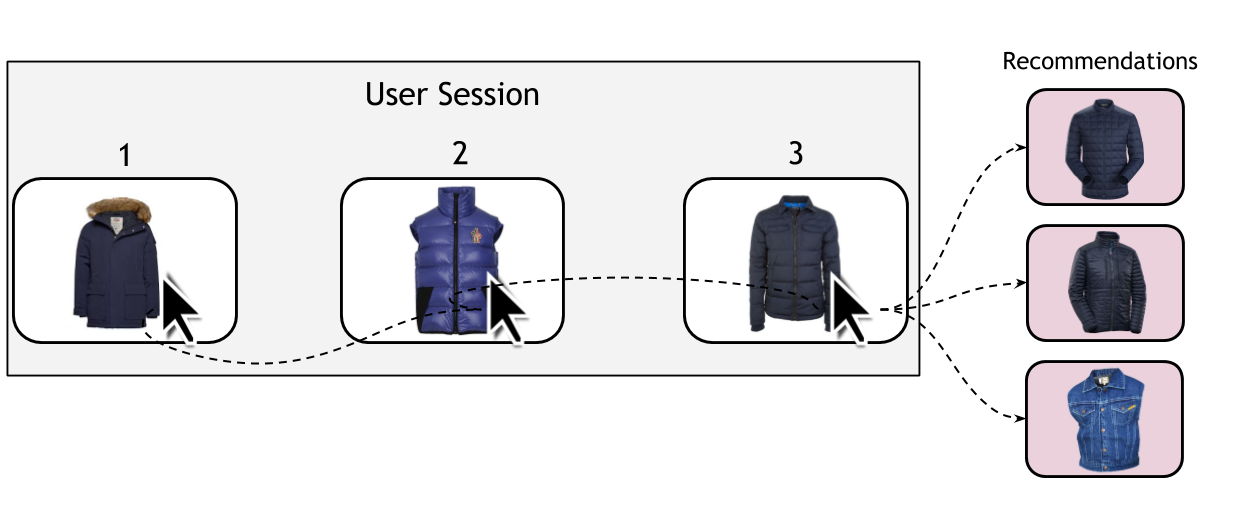}
    \caption{An example of session-based recommendation. $3$ different items are clicked by the user. The recommendation system predicts other candidate items that can be viewed next by the same user.}
    \label{fig:Session-Recommendation}
\end{figure}

In the recent few years, more attention has been paid to session-based recommendation due to security policies of collecting personalized user behavior data. In particular, a more research focus has been directed towards the anonymous session-based recommendation systems. The main reason for this research focus is to comply with the recent (\emph{GDPR}) rules that make the collection of personalized data about users more challenging to protect the user's privacy \citep*{mohallick2018towards}. Also, it is not easy to collect enough long-term user profile data to recommend the next items reliably.  Figure \ref{fig:Session-Recommendation} shows an example of a session-based recommendation where the user has a stream of click events on multiple items. The recommendation system tries to predict the next items to be viewed by the same user based on the information made available during the same session only.

\label{dnn}
Deep neural networks are a subset of machine learning technologies that have attracted significant attention in the past decade. Such techniques have achieved outstanding performance in a wide range of domains, including natural language processing, medical diagnosis, speech recognition, and computer vision. In practice, the main advantage of deep learning over traditional machine learning techniques is their automatic feature extraction ability. It allows learning complex functions to be mapped from the input space to the output space without a human intervention \citep*{lecun2015deep}. Recently, different approaches have been proposed to use deep neural networks in recommendation systems \citep*{wang2015collaborative, kim2016convolutional}. In particular, different deep learning models were used for modeling the sequence of the user navigation behavior in the online services to be used in the next item recommendation \citep*{hidasi2015session, kang2018self, liu2018stamp, wu2019session}. These works showed a competitive performance compared to traditional approaches like sequential pattern mining techniques, nearest neighbor algorithms, and traditional Markov models \citep*{ludewig2018evaluation}.

\label{previous-studies}
Few studies have been conducted to evaluate the session-based algorithms. \citet*{jannach2017recurrent} compared the heuristics-based nearest neighbor baseline algorithm with a basic recurrent neural network ({RNN}). The results of this study showed that deep learning methods fall behind basic algorithms like neighbor-hood methods. However, during the last couple of years, many advancements have been proposed using deep-learning in the session-based recommendation, leading to the rise of several neural-based architectures. \citet*{ludewig2018evaluation}, for instance, conducted a study to compare many baseline algorithms in session-based recommendation using four datasets in the e-commerce field and four others in music and playlists recommendation. Although this study included only a single neural-based model and lacked the evaluations of different deep-learning approaches in the literature, it was recently extended to include the state-of-the-art deep learning models \citep*{ludewig2019empirical}. However, the empirical evaluation conducted in \citep*{ludewig2018evaluation, ludewig2019empirical} by training models on full datasets makes it difficult to understand the drawbacks of each model and why exactly a specific model outperforms the others in a particular dataset.

\label{contribution}
In this paper, we extend the previous studies \citep*{kamehkhosh2017comparison, ludewig2018evaluation, ludewig2019empirical} that compared the overall performance of 11 simple algorithms and 6 deep-learning models on four e-commerce datasets. In this work, we focus on studying the effect of varying the characteristics of a dataset on the performance of each model. In particular, our main contributions are as follows:
\begin{itemize}
    \item We carry out an extensive \emph{evaluation} and \emph{benchmarking} of the state-of-the-art neural-based approaches in the session-based recommendation, including recurrent neural networks, convolutional neural networks, and attention-based networks along with a group of the most popular baseline techniques in the recommendation field like nearest-neighbors, frequent pattern mining and matrix factorization.
    \item we evaluate the performance of the different models based on \emph{various characteristics of train and test dataset splits} obtained from four different e-commerce benchmark datasets, namely \textit{RECSYS}\footnote{http://2015.recsyschallenge.com/}, \textit{CIKMCUP}\footnote{https://cikm2016.cs.iupui.edu/cikm-cup/}, \textit{RETAIL ROCKET}\footnote{https://www.kaggle.com/retailrocket/ecommerce-dataset}, and \textit{TMALL}\footnote{https://www.tmall.com/}
    \item Our experiments elaborate on the evaluation process based on various dataset characteristics. Hence, we divide the datasets according to the values of various characteristics like session length, item frequency, and data sizes. These experiments revealed some insights that could help understand when some models are poorly performing and open new research horizons of what are the needed improvements for each model, which were hard to observe from the previous studies.
    \item An interpretable decision tree model is used to accurately recommend the best performing model according to the dataset characteristics.
    \item Current drawbacks of session-based recommendation systems are discussed with proposed solutions to overcome these issues, which could yield better results in some domains.
\end{itemize}
We divide our benchmarking study into separate sets of experiments aiming to answer a different research question by each set. First, we evaluated the performance of different models against different session lengths and frequencies of items. Second, we investigated the effect of the collected data recency on the models' performance, which could help avoid data leakage problems and deceiving accuracy during training. Third, the effect of the training data size is evaluated for different models. Finally, we present a comparison of different approaches in terms of time and memory resource consumption during both training and inference. The aim of this study is mainly to understand \textit{what are the main characteristics of datasets that profoundly affect different models' performance by carrying out a micro-analysis evaluation for the session-based algorithms on real-world e-commerce datasets}. This study could help to \emph{improve the selection of the recommendation algorithm according to the target dataset and highlight the weaknesses of different models for further improvement}.

\label{organization}
The paper is organized as follows. In Section \ref{sec:2}, a short survey of session-based recommendation systems have been discussed. Section \ref{sec:3} presents a detailed description of different algorithms and models evaluated in our experiments. Section \ref{sec:4} describes the experiments setup and the research questions to be answered, and Section \ref{sec:5} shows the results and discussion of the evaluation experiments. Finally, in Section \ref{sec:6}, the main insights of our study have been summarized in addition to thoughts of future research directions.

\section{Review of Deep Learning Approaches in Session-Based Recommendation}
\label{sec:2}

The session-based recommendation is a particular type of sequence-aware recommendation that is a general class of recommendation systems. The decisions made by these systems are mainly based on the user short-term intention defined by a session. This session is represented by a set of the user-item interaction pairs in a short period of time. Also, various types of attributes can characterize these interactions like user's attributes (e.g., gender and age), item's attributes (e.g., color and size), and action types (e.g., add-to-cart and add-to-wish-list). The input of these recommendation systems is a chronologically ordered set of user-item actions and the output is a score-list of the ranking of items based on the likelihood that user preferences match these items \citep*{quadrana2018sequence}. Even though e-commerce is the most critical application for the session-based recommendation, there are many other applications such as recommendation for music playlist, movie, and online course \citep*{jannach2015adaptation}.

\begin{table}
\centering
\caption{Summary of current state-of-the-art neural session-based recommendation architectures.}
\resizebox{\textwidth}{!}{\begin{tabular}{|l|c|c|c|c|c|c|c|}
\hline
\multicolumn{1}{|c|}{\textbf{Model name}} & \textbf{\begin{tabular}[c]{@{}c@{}}Date\end{tabular}} & \textbf{\begin{tabular}[c]{@{}c@{}}Personalized\\ recommendation\end{tabular}} & \textbf{\begin{tabular}[c]{@{}c@{}}Main layers \\ of architecture\end{tabular}} & \textbf{\begin{tabular}[c]{@{}c@{}}Attention/\\ Memory\end{tabular}} & \textbf{\begin{tabular}[c]{@{}c@{}}Support\\ items \\ features\end{tabular}} & \textbf{Framework} & \textbf{\begin{tabular}[c]{@{}c@{}}Open \\ source\end{tabular}} \\ \hline
Item2Vec \citep*{DBLP:journals/corr/BarkanK16} & 2015 & \xmark & \begin{tabular}[c]{@{}c@{}}Feed forward\\ network\end{tabular} & \xmark & \xmark & Genism \tablefootnote{https://github.com/Bekyilma/Recommendation-based-on-sequence-} & \cmark \\ \hline
GRU4Rec \citep*{hidasi2015session} & 2015 & \xmark & \begin{tabular}[c]{@{}c@{}}Recurrent \\ layers network\end{tabular} & \xmark & \xmark & Theano \tablefootnote{https://github.com/hidasib/GRU4Rec} & \cmark \\ \hline
P-GRU4Rec \citep*{hidasi2016parallel} & 2016 & \xmark & \begin{tabular}[c]{@{}c@{}}Recurrent\\ layers network\end{tabular} & \xmark & \cmark & \xmark & \xmark \\ \hline
Conv3D4Rec \citep*{tuan20173d} & 2017 & \xmark & \begin{tabular}[c]{@{}c@{}}3D Convolutions\\ networks\end{tabular} & \xmark & \cmark & \xmark & \xmark \\ \hline
NARM \citep*{li2017neural} & 2017 & \xmark & \begin{tabular}[c]{@{}c@{}}Recurrent\\ layers network\end{tabular} & \cmark & \xmark & Theano \tablefootnote{https://github.com/lijingsdu/sessionRec\_NARM} & \cmark \\ \hline
IIRNN \citep*{ruocco2017inter} & 2017 & \cmark & \begin{tabular}[c]{@{}c@{}}Recurrent\\layers network\end{tabular} & \xmark & \xmark & Tensorflow \tablefootnote{https://github.com/olesls/master\_thesis} & \cmark \\ \hline
HGRU4Rec \citep*{quadrana2017personalizing} & 2017 & \cmark & \begin{tabular}[c]{@{}c@{}}Recurrent\\ layers network\end{tabular} & \xmark & \cmark & Theano \tablefootnote{https://github.com/mquad/hgru4rec} & \cmark \\ \hline
GRU4Rec+ \citep*{hidasi2018recurrent} & 2018 & \xmark & \begin{tabular}[c]{@{}c@{}}Recurrent\\ layers network\end{tabular} & \xmark & \xmark & Theano \tablefootnote{https://github.com/hidasib/GRU4Rec} & \cmark \\ \hline
STAMP \citep*{liu2018stamp} & 2018 & \xmark & \begin{tabular}[c]{@{}c@{}}Feed forward\\ network\end{tabular} & \cmark & \xmark & Tensorflow \tablefootnote{https://github.com/uestcnlp/STAMP} & \cmark \\ \hline
SASRec \citep*{kang2018self} & 2018 & \cmark & \begin{tabular}[c]{@{}c@{}}Feed forward\\ network\end{tabular} & \cmark & \xmark & Tensorflow \tablefootnote{https://github.com/kang205/SASRec} & \cmark \\ \hline
CASER \citep*{tang2018personalized} & 2018 & \cmark & \begin{tabular}[c]{@{}c@{}}Convolutional\\ layers network\end{tabular} & \xmark & \xmark & Pytorch \tablefootnote{https://github.com/graytowne/caser\_pytorch} & \cmark \\ \hline
NextItNet \citep*{yuan2019simple} & 2019 & \xmark & \begin{tabular}[c]{@{}c@{}}Dilated convolutional\\layers network\end{tabular} & \xmark & \xmark & Tensorflow \tablefootnote{https://github.com/fajieyuan/nextitnet} & \cmark \\ \hline
SRGNN \citep*{wu2019session} & 2019 & \xmark & \begin{tabular}[c]{@{}c@{}}Graph neural\\ network\end{tabular} & \cmark & \xmark & \begin{tabular}[c]{@{}c@{}}Tensorflow/\ Pytorch \tablefootnote{https://github.com/CRIPAC-DIG/SR-GNN}\end{tabular} & \cmark \\ \hline
CSRM \citep*{wang2019collaborative} & 2019 & \xmark & \begin{tabular}[c]{@{}c@{}}Recurrent \\ layers network\end{tabular} & \cmark & \xmark & Tensorflow \tablefootnote{https://github.com/wmeirui/CSRM\_SIGIR2019} & \cmark \\ \hline
BERT4Rec \cite{sun2019bert4rec} & 2019 & \cmark & \begin{tabular}[c]{@{}c@{}}Transformer\\based network\end{tabular} & \cmark & \xmark & Tensorflow \tablefootnote{https://github.com/FeiSun/BERT4Rec} & \cmark \\ \hline
DCN-SR \citep*{chen2019dynamic} & 2019 & \xmark & \begin{tabular}[c]{@{}c@{}}Recurrent\\layers network\end{tabular} & \cmark & \xmark & \xmark & \xmark \\ \hline
\end{tabular}}
\label{tab:sotas}
\end{table}

Early research works tackling session-based recommendation problems with the nearest neighbors and the frequent pattern mining techniques \citep*{bonnin2015automated}. However, these works are instance-based algorithms, that take much time for making predictions. Therefore, they are not suitable for real-time use cases, such as e-commerce. Later on, other research works proposed using more advanced techniques, such as Markov chain models in sequence modeling \citep*{garcin2013personalized,hosseinzadeh2015adapting}. The problem of the state-space explosion in Markov models was treated by using attributes of some items to limit the space of the next items to be recommended \citep*{tavakol2014factored}. Additionally, classical matrix factorization techniques were combined with Markov chains in different variations and applied in a wide range of domains as in \citep*{cheng2013you,he2016fusing}.

In the last few years, different deep learning approaches have been adopted in the session-based recommendation. The main advantage of deep learning approaches is their ability to extract features automatically. This advantage allows learning complex functions to be mapped from the input space to the output space without human intervention \citep*{lecun2015deep}. For example, a neural-based model named GRU4Rec used Gated Recurrent Units ({GRU}s) in {RNN}s to predict the next item to be clicked by the user \citep*{hidasi2015session}. The model was trained by minimizing loss functions that include pairwise losses comparing the target item score with the maximal score among negative samples. The likelihood of these samples is taken into account in proportion to the target item's maximal score. The used losses showed excellent performance by correctly ranking the predicted items and overcoming the vanishing gradient problem in {RNN}s \citep*{hidasi2018recurrent}.
The GRU4Rec architecture was further extended by using a modified version of the original negative sampling approach, where the likelihood score of the next recommended item is calculated for a subset of items as it would be impractical to do it for the whole list of items \citep*{hidasi2018recurrent}. The new sampling method uses additional negative samples shared by all the session sequences within the same mini-batch. Besides, it updates a small percentage of the network weights for each mini-batch to make the training process faster. These samples were chosen based on items' popularity, which gives more chances to include most of the high scoring negative examples. This approach leads to excellent improvement in the performance of the model. Also, the same architecture was adopted to support multiple item features instead of unique identifiers only in a parallel training scheme. It was evaluated against item K-nearest neighbors showing a good improvement \citep*{hidasi2016parallel}.
Furthermore, \citet*{quadrana2017personalizing} proposed a method for adapting {RNN} in personalized session-based recommendation with cross-session information transfer among user sessions using a hierarchical {RNN} model such that the output hidden state from the network for a particular session is passed as input to a higher level {RNN} for the next session of the same user. A hybrid architecture of two {RNN}s was proposed for a personalized session-based recommendation that aims mainly in targeting the session cold-start problem by learning from the user personal recent sessions \citep*{ruocco2017inter}. Convolutional neural networks ({CNN}s) were also used in the session-based recommendation. In particular, \citet*{tuan20173d} used a 3D-CNN with character level encoding to combine session clicks with the textual descriptions of the items to generate recommendations. Similarly, a generative {CNN} was proposed by embedding the clicked items into a 2-dimensional matrix and treated as input images to the {CNN} \citep*{yuan2019simple}. Graph neural networks were recently used to capture complex transitions among items after modeling the sequence of the events of a session as a graph-structured data without adequate user behavior in a session \citep*{DBLP:journals/corr/abs-1902-07243}. \citet*{wang2019collaborative} proposed a novel framework using two parallel memory encoders to make use of collaborative neighborhood sessions information in addition to the current session information followed by a selective fusion of both encoders output.
 
 After discovering the attention concept in neural networks that leads to a great improvement in neural machine translation tasks \citep*{graves2014neural,vaswani2017attention}, attention networks are widely adopted in the session-based recommendation \citep*{li2017neural,liu2018stamp,chen2019dynamic}. For instance,  a hybrid encoder with attention is used to model the user sequential behavior \citep*{li2017neural}, which outperformed the long-term memory models like {GRU4Rec} \citep*{li2017neural}. Also, a short term attention priority model was introduced such that attention weights are computed from the total session context and enhanced by the current user's interest represented by the last clicked item \citep*{liu2018stamp}. Besides, \citet*{devlin2018bert} adopted the current state-of-the-art BERT transformer network, used widely in the natural language processing domain, in personalized session-based recommendation \citep*{sun2019bert4rec}. Most of the neural-based solutions, in the session-based recommendation, generate a static representation for users' long-term interests. Such representation might be an issue as its importance for predicting the next recommended item is dynamic and related also to the short-term preferences. Hence, a co-attention network was proposed to recognize the dynamic interaction between the user's long and short-term interests to generate a co-dependent representation of the users' interests \citep*{chen2019dynamic}. However, the usage of the transformer networks in generalized session-based recommendations with the incorporation of item features are still open research areas. Table \ref{tab:sotas} summarizes the current state-of-the-art neural network architectures for personalized/non-personalized session-based recommendation.
 
\begin{table}[t]
\centering
\caption{Notation of different variables used in explaining the evaluated methods}
\label{tab:notation}
\resizebox{\textwidth}{!}{\begin{tabular}{|l|c|} 
\hline
\multicolumn{1}{|c|}{ \textbf{Symbol} } & \textbf{Description}  \\ 
\hline
$\mathcal{I}$  & Set of available items \\ 
\hline
$N$  & Total number of available items = $|\mathcal{I}|$  \\ 
\hline
$I_n$  & Item with index $n \in \mathcal{I}$ where $n \in N$  \\ 
\hline
$x_{i_t}$  & The $i^{th}$ click event in a session starting at time $t$  \\ 
\hline
$L_t$  & The length of session starting at time $t$  \\ 
\hline
$S_t$  & \begin{tabular}[c]{@{}c@{}}A session started at time $t$ representing the sequence of clicked items\\ $\{x_{1_t}, x_{2_t}, ..., x_{L_t}\}$ \end{tabular} \\ 
\hline
$1_{EQ}(x,y)$  & $=1$ if $x = y$, and $0$ otherwise.  \\ 
\hline
$\mathcal{S}_{D}$, $\mathcal{S}_{TR}$, $\mathcal{S}_{TE}$  & Set of general dataset $D$, training and testing sets sessions respectively  \\ 
\hline
$dis(j, k)$  & Distance between items at indices $j$, and $k$ in the session click stream  \\ 
\hline
$sim(S_i, S_j)$  & Similarity distance between sessions $S_i$ and $S_j$  \\ 
\hline
$1_{IN}(x, \mathbf{Y})$  & $ = 1$ if $x$ is one of elements in vector $Y$, and $0$ otherwise  \\ 
\hline
$r_{IN}(x, \mathbf{Y})$  & $ =$ rank of $x$ if it is one of elements in vector $Y$, and $0$ otherwise  \\ 
\hline
$W_t(S_t)$  & A weighting function for items clicked in session $S_t$  \\ 
\hline
$E_{S_t}$, $E_i$  & Embedding vector for session $S_t$, or item $I_i$ respectively  \\ 
\hline
$\argmax_{K}(\hat{\mathbf{Y}})$  & Index of items with top $K$ predicted scores in vector \textbf{Y}.  \\ 
\hline
$\mathcal{U}( \mathbf{Y} )$  & Set of unique items in vector \textbf{Y}.  \\ 
\hline
$\mathcal{F}(\mathbf{Y})$  & Frequencies of items in vector \textbf{Y} from the training set.  \\
\hline
\end{tabular}}
\end{table}

\section{Detailed Evaluated Approaches}
\label{sec:3}
In this section, all the algorithms covered in our evaluation study are explained in detail, and for the sake of simplicity, the following notation in Table \ref{tab:notation} is used throughout the whole section.
 
\subsection{Baseline Approaches}
We selected a set of five baseline algorithms to be included in this study based on the previous study in \citep*{ludewig2018evaluation}. In particular, our selection was done based on two different criteria. First, we selected at least one method from each family of algorithms, which showed excellent performance at different session-based recommendation tasks. Second, we chose the method with the best overall performance compared to the other methods within the same family. Therefore, the selected algorithms are as following: session-based popular products (S-POP) as a simple heuristic algorithm \citep*{adomavicius2005toward}, and simple association rules ({AR}) and simple sequential rules ({SR}) as representatives of frequent pattern mining algorithms \citep*{agrawal1993mining}. Vector session-based K-nearest neighbors (VSKNN) \citep*{ludewig2018evaluation} and Session-based matrix factorization ({SMF}) \citep*{ludewig2018evaluation} are selected from the nearest neighbors, factorization-based methods, respectively. 

\subsubsection{Session-Based Popular Products}
S-POP is one of the most widely used baseline recommendation algorithms \citep*{adomavicius2005toward,steck2011item}. These algorithms make a recommendation based on the most frequent item viewed by the user in the current session. In short, if a user clicked on an item, $I_n$, multiple times during the same session, this reflects a clear sign of the user's interest in that item. Hence, recommending the same item to the user again is a reasonable decision. In some cases, the S-POP recommendation process is limited to the top popular $K$ items while ignoring the rest of the items. This constraint ensures that the recommended items belong to the most popular ones among all users.

Score of a specific item $I_n$ in a session $S_t$ is computed as follows:
\begin{equation}
    Score(I_n, S_t) = \sum_{i=1}^{L_t} 1_{EQ}(x_{i_t}, I_n).
\end{equation}

\subsubsection{Simplified Association Rules}
Association rules (AR) is one of the frequent pattern mining approaches such that it captures the size for the frequency of patterns of events, $N$, and recommend the most frequent ones \citep*{agrawal1993mining}. In the case of session-based recommendation, \citet*{ludewig2018evaluation} used a simplified version of association rules of size $N=2$ to have a reasonable computational complexity. In their work, the occurrence of any two subsequent items ($I_i, I_j$) at the same session $S$ is stored. During prediction, the last item viewed by the user, $x_{L_t}$, is used to find all the candidate similar items by choosing the most frequent item pairs, ($x_{L_t}, I_n$) where $n \in N$. Therefore, an arbitrary item $I_n$ is recommended if it has a score among the top predicted ones. This score is computed as follows:
\begin{equation}
    Score(I_n, S_t) = \sum_{S_i \in \mathcal{S}_{TR} } \sum_{j=1}^{|L_i|} \sum_{k=1}^{|L_i|} 1_{EQ}(x_{L_t}, x_{j_i}) . 1_{EQ}(I_{n}, x_{k_i}).
\end{equation}

\subsubsection{Simplified Sequential Rules}
Sequential rules is also a frequent pattern mining approach. Here, the order of the session events is taken into account in contrast with {AR} that depends on the support of the items only. A simplified form of sequential rules ({SR}) is used such that a rule is created between two items ($I_i, I_j$) when they appear in sequential events \citep*{kamehkhosh2017comparison}. Each rule in {SR} is assigned a weight that is a function of the linear distance between the items ($I_i, I_j$) as in Eq. \ref{eq:3}. The rules between near events are assigned larger weights than rules between far events. The scores of different items to be recommended can be evaluated using the following:
\begin{equation}
\label{eq:3}
\begin{aligned}[b]
    & Score(I_n, S_t) = \sum_{S_i \in \mathcal{S}_{TR} } \sum_{j=2}^{|L_i|} \sum_{k=1}^{x-1} 1_{EQ}(x_{L_t}, x_{k_i}) . 1_{EQ}(I_{n}, x_{j_i}) . dis(j,k),
\end{aligned}
\end{equation}
    where $dis(j,k) = \left(1 - 0.1(j-k) \right)$ if $ j - k < 10$ otherwise $dis(j, k) = 0$.
    
\subsubsection{Vector Multiplication Session-based K-Nearest Neighbors}
Nearest neighbor algorithms show excellent performance in session-based recommendation \citep*{kamehkhosh2017comparison}. However, they have many different variant schemes which can be applied according to the domain type like item based nearest neighbors \citep*{wen2008recommendation} which depends on predicting similar items to the last one viewed by the user. On the other hand, session-based nearest neighbors consider the viewed items in the whole session and try to find neighboring sessions with similar items to be used in predicting the next recommended items \citep*{bonnin2015automated}. \citet*{ludewig2018evaluation} evaluated multiple variants of nearest neighbor algorithms. In their work, it has been shown that vector multiplication session-based K-nearest neighbors (VSKNN) has outperformed pattern mining and matrix factorization methods in most of the evaluated datasets. Besides, it has a competitive performance to {RNN}s and even outperforms them in multiple datasets. VSKNN is considered as one of the session-based nearest neighbors algorithms, where recent items clicked by the user take larger weights than older items. This way, more emphasis is given for the recent events made by the user. The score of an item $I_n$ to be recommended for the next item is computed as
\begin{equation}
    Score(I_n, S_t) = \sum_{S_i \in \mathcal{S}_{TR} } \left[sim(S_t, S_i) . W_t(S_t)]\right . 1_{IN}(I_n, S_i),
\end{equation}
where the similarity distance, $sim(S_t, S_i)$, can be set to the cosine distance, and $W_t(S_t)$ is a weighting function of the items according to their positions in the session $S_t$. This weighting function usually gives higher weights to the recently clicked items \citep*{ludewig2018evaluation}.

\subsubsection{Session-based Matrix Factorization}
{SMF} is a matrix factorization based approach designed for the task of session-based recommendation \citep*{ludewig2018evaluation}. This approach was inspired by the factorized personalized Markov chains \citep*{kabbur2013fism,he2016fusing} for sequential recommendation tasks. In {SMF}, classical matrix factorization and factorized Markov chains are combined with a hybrid approach. In particular, the latent user vector was replaced by an embedding vector that represents the current session. During prediction making, the score of a candidate item is computed as the weighted sum of the whole session preferences and the sequential dynamics representing the transition probability from the last clicked item by the user to the candidate item to be recommended by the model.
We used the model implementation by \citet*{ludewig2018evaluation}. The SMF showed a better performance than other factorization-based methods over multiple datasets.

\begin{figure*}[!ht]
    \centering
    \includegraphics[width=\textwidth]{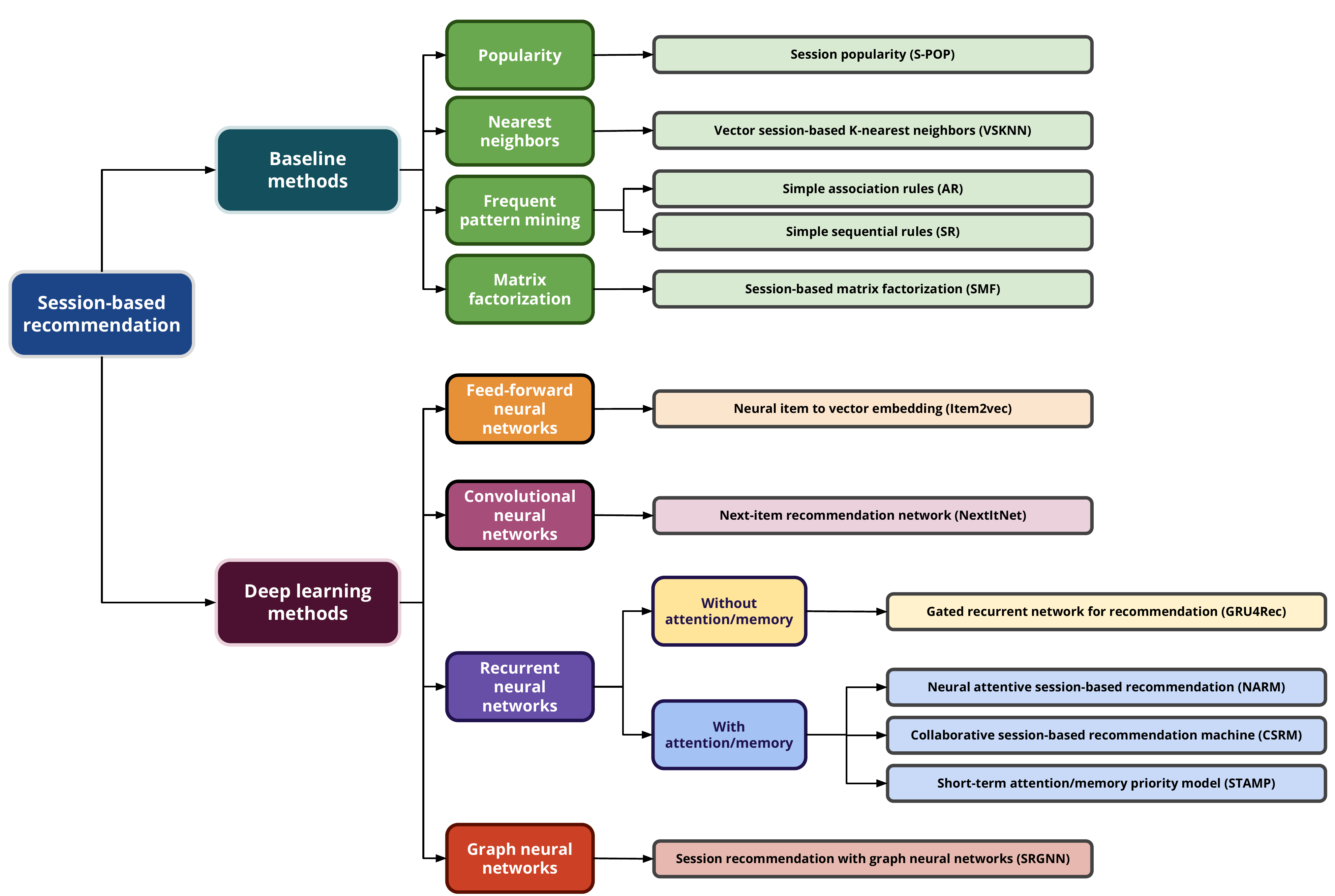}
    \caption{List of evaluated baseline and deep learning session-based recommendation methods.}
    \label{fig:EvaluatedMethods}
\end{figure*}

\subsection{Deep Learning Approaches}
Many deep learning architectures were proposed in the literature for session-based recommendation. These architectures vary in the types of their layers. For instance, \citet*{hidasi2015session} presented the first study using {RNN}s with {GRU}s in session-based recommendation. \citet*{tuan20173d} and \citet*{yuan2019simple} used convolutional networks in modeling the session context. \citet*{li2017neural, liu2018stamp} proposed different attention mechanisms to enhance the performance of the {RNN}s. Recently, \citet*{wu2019session} exploited graph neural networks in session-based recommendation.

In our study, we limited the selection to include only the current state-of-the-art and well-cited architectures proposed in the range of last four years, published in top tier venues, and have an open source implementation. Additionally, we refined our list to select the models that can be used in making generalized (non-personal) predictions without the need of collecting personal user profile to comply easily with the GDPR requirements (Section \ref{intro}). The final list of the chosen architectures includes neural item embedding algorithm ({Item2Vec}) proposed by \citet*{DBLP:journals/corr/BarkanK16}, extended version of {GRU}s neural networks ({GRU4Rec+}) by \citet*{DBLP:journals/corr/HidasiK17}, neural attentive network ({NARM}) \citet*{li2017neural}, graph neural network proposed by \citet*{wu2019session}, short-term attention priority network ({STAMP}) by \citet*{liu2018stamp} as well as convolutional generative network for session-based recommendation ({NextItNet}) \citep*{yuan2019simple}, and collaborative neural network with parallel memory modules ({CSRM}) proposed by \citet*{wang2019collaborative}.

\subsubsection{Neural item to vector embedding}
\citet*{mikolov2013efficient} introduced a conversion for the items into embedding vectors in a latent space based on the session context of clicked items. This idea is an adaptation of the \emph{Word2Vec} algorithm that converts words into a vector space in an efficient way that enhances neural machine translation task performance by having two close vectors for similar words used in the same context. Similarly, Item2Vec uses the skip-gram with negative sampling neural word embedding to find out vector representations for different items that infer the relationship between an item and its surrounding items in a session. During the prediction phase, candidate items get scores according to the similarity distance between their embedding vectors and the average of the embedding vectors of the session items \citep*{DBLP:journals/corr/BarkanK16}.

\subsubsection{Gated recurrent neural networks for session-based recommendation}
\label{GRU4Rec}

\begin{figure}
    \centering
    \includegraphics[width=\textwidth]{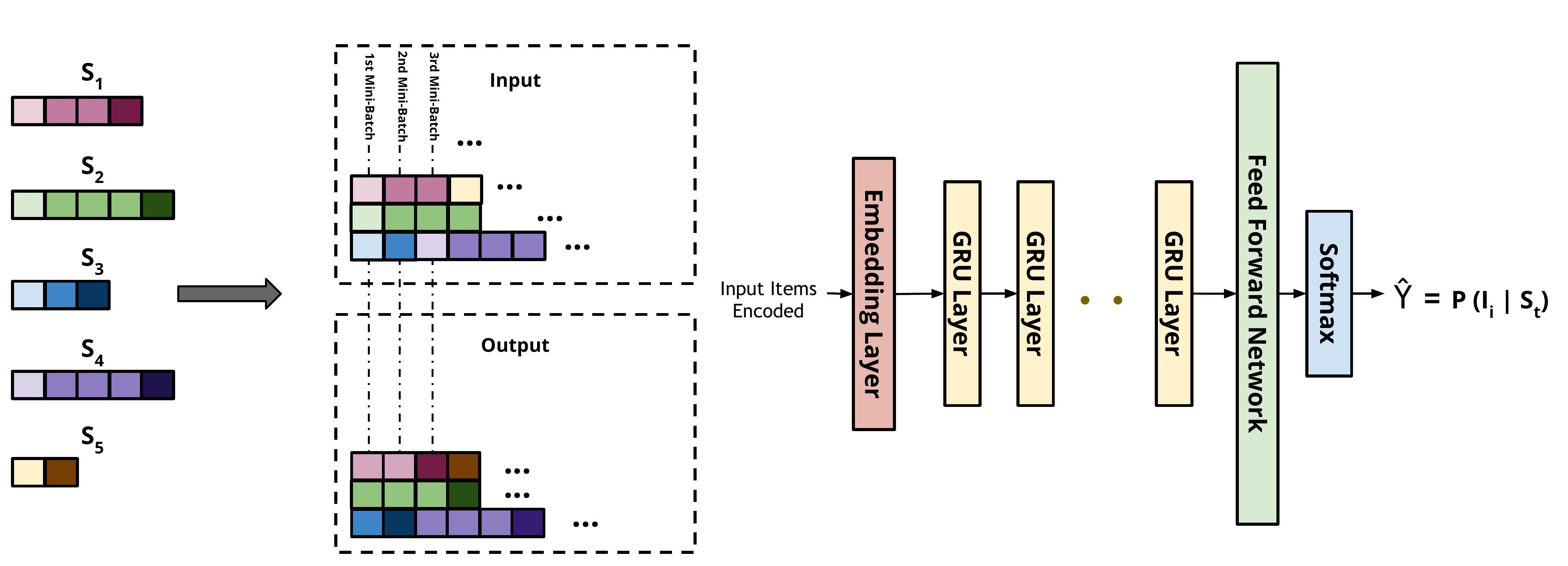}
    \caption{{GRU4Rec} model architecture \citep*{hidasi2015session}}
    \label{fig:GRU4Rec}
\end{figure}

One of the first successful approaches for using {RNN}s in the recommendation domain is the GRU4Rec network \citep*{hidasi2015session}. A {RNN} with {GRU}s was used for the session-based recommendation. A novel training mechanism called session-parallel mini-batches is used in {GRU4Rec}, as shown in Figure \ref{fig:GRU4Rec}. Each position in a mini-batch belongs to a particular session in the training data. The network finds a hidden state for each position in the batch separately, but this hidden state is kept and used in the next iteration at the positions when the same session continues with the next batch. However, it is erased at the positions of new sessions coming up with the start of the next batch. The network is always updated with the session beginning and used to predict the subsequent events. {GRU4Rec} architecture is composed of an embedding layer followed by multiple optional {GRU} layers, a feed-forward network, and a softmax layer for output score predictions for candidate items. The session items are one-hot-encoded in a vector representing all items' space to be fed into the network as input. On the other hand, a similar output vector is obtained from the softmax layer to represent the predicted ranking of items. Additionally, the authors designed two new loss functions, namely, Bayesian personalized ranking (BPR) loss and regularized approximation of the relative rank of the relevant item (TOP1) loss. BPR uses a pairwise ranking loss function by averaging the target item's score with several sampled negative ones in the loss value. TOP1 is the regularized approximation of the relative rank of the relevant item loss. Later, \citet*{hidasi2018recurrent} extended their work by modifying the two-loss functions introduced previously by solving the issues of vanishing gradient faced by TOP1 and BPR when the negative samples have very low predicted likelihood that approaches zero. The newly proposed losses merge between the knowledge from the deep learning and the literature of learning to rank. The evaluation of the new extended version shows a clear superiority over the older version of the network. Thus, we have included the extended version of the {GRU4Rec} network, denoted by {GRU4Rec+}, in our evaluation study.

\subsubsection{Neural attentive session-based recommendation}
\label{NARM}
{NARM} is one of the session-based recommendation systems based on sequence modeling using an attention mechanism \citep*{li2017neural}. The main advantage of this model is introducing a solution to the long-term memory models like {GRU4Rec}(\ref{GRU4Rec}). The model is characterized by hybrid encoders with an attention network to model the user sequential behavior and capture the main purpose of the session combined as a unified session representation.

{NARM} architecture has two types of encoders: 
\begin{enumerate}
    \item {GRU} network representing the global encoder, which takes the entire previous user interactions during the session as input and produces the user's sequential behavior as output.
    \item Local encoder that is a {GRU} network similar to the global encoder. However, its role is to involve an item level attention mechanism to allow the decoder to dynamically select a linear combination of different items from the input sequence, and focus more on important items that can capture the user's main purpose within a particular session.
\end{enumerate} 
Finally, both encoders' outputs are concatenated with each other to form an extended representation of the session. They are fed again into a bi-linear decoder along with item embedding vectors to compute the similarity score between current session representation and candidate items to be used in ranking items to be predicted next.

\subsubsection{Short-term attention/memory priority model}
\label{STAMP}
{STAMP} is one of the approaches that replaces complex recurrent computations in {RNN}s with self-attention layers \citep*{liu2018stamp}. The model presents a novel attention mechanism in which the attention scores are computed from the user's current session context and enhanced by the sessions' history. Thus, the model can capture the user interest drifts, especially during long sessions and outperform other approaches like {GRU4Rec} \citep*{hidasi2015session} that uses long term memory but still not efficient in capturing user drifts.
\begin{figure}
    \centering
    \includegraphics[width=0.8\textwidth]{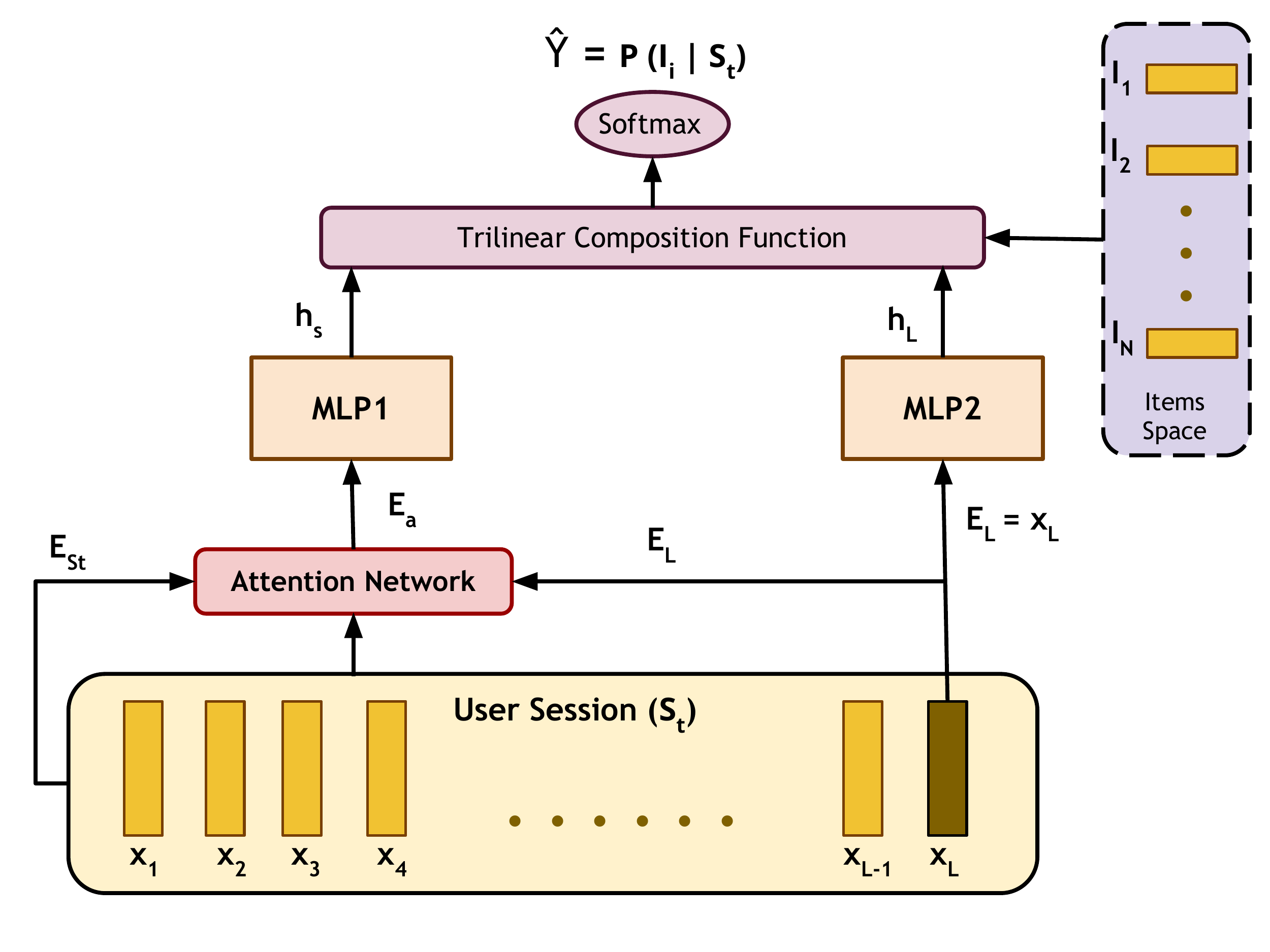}
    \caption{{STAMP} model architecture \citep*{liu2018stamp}}
    \label{fig:STAMP}
\end{figure}
Figure \ref{fig:STAMP} shows the model architecture where the input is two embedding vectors ($\mathbf{E_L}, \mathbf{E_{S_t}}$). The former denotes the embedding of the last item $x_L$ clicked by the user in the current session, which represents the short term memory of the user's interest. The later represents his overall interest through the full session clicked items. $\mathbf{E_{S_t}}$ vector is computed by averaging the items embedding vectors throughout the whole session memory ($x_1, x_2, ..., x_{L}$).
An attention layer is used to produce a real-valued vector $\mathbf{E_a}$, where this layer is responsible for computing the attention weights corresponding to each item in the current session. In this way, we avoid treating each item in the session equally important and paying more attention to only related items, which improves the capturing of the drifts in the user interest.
Both $\mathbf{E_a}$ and $\mathbf{E_L}$ flow into two multi-layer perceptron networks identical in shape but have separate independent parameters for feature abstraction. Finally, a trilinear composition function, followed by a softmax function, is used for the likelihood calculation of the available items to be clicked next by the user and to be used in the recommendation process.

\subsubsection{Simple generative convolutional network}
{NextItNet} was proposed to use convolutional neural networks in the session-based recommendation. The session made by a user is converted into a 2-dimensional latent matrix and fed into a convolutional neural network like images \citep*{yuan2019simple}.

{NextItNet} is considered as an extension over the recent convolutional sequence embedding recommendation model ({Caser}) by \citet*{tang2018personalized}. However, {NextItNet} addresses two main limitations of applying {CNN}s in sequence modeling in {Caser}, which are obvious in long sessions. First, the items sequences in a session can have a variable length, which means that a large number of different size images are needed to represent a session. Consequently, fixed-size convolutional filters may fail in dealing with such cases. However, large filters with a filter width similar to the image width of an item inside the session sequence, and followed by max-pooling layers, are used to ensure that the produced feature maps have the same length. Second, these small filters are not able to find well-representing embedding vectors for the session items. In {NextItNet}, a huge number of inefficient convolutional filters are replaced with a series of 1-dimensional dilated convolution layers. The dilated layers are responsible for increasing the receptive field and dealing with different session lengths instead of the standard 2D convolution layers. Thus, the max-pooling layers are omitted as they can not distinguish the important features in the map if they occur once or multiple times while ignoring the position of these features. Additionally, {NextItNet} makes use of the residual blocks effectively in the recommendation systems, which can ease the optimization for much deeper networks than the shallow convolutional network in {Caser} that can not model complex relations between items in a user session.

\subsubsection{Session-based recommendation with graph neural networks}
{SRGNN} was introduced recently by \citet*{wu2019session}. The session sequences are modeled as a graph-structured data and the graph neural network ({GNN}) task is to capture the complex transitions among items. This architecture was proposed to solve mainly two problems with other approaches. First, most other models can not estimate the user interest without adequate interactions in a session. Secondly, most of the models focus on single way transitions between items and neglect transitions among the context instead.
\begin{figure}
    \centering
    \includegraphics[width=\textwidth]{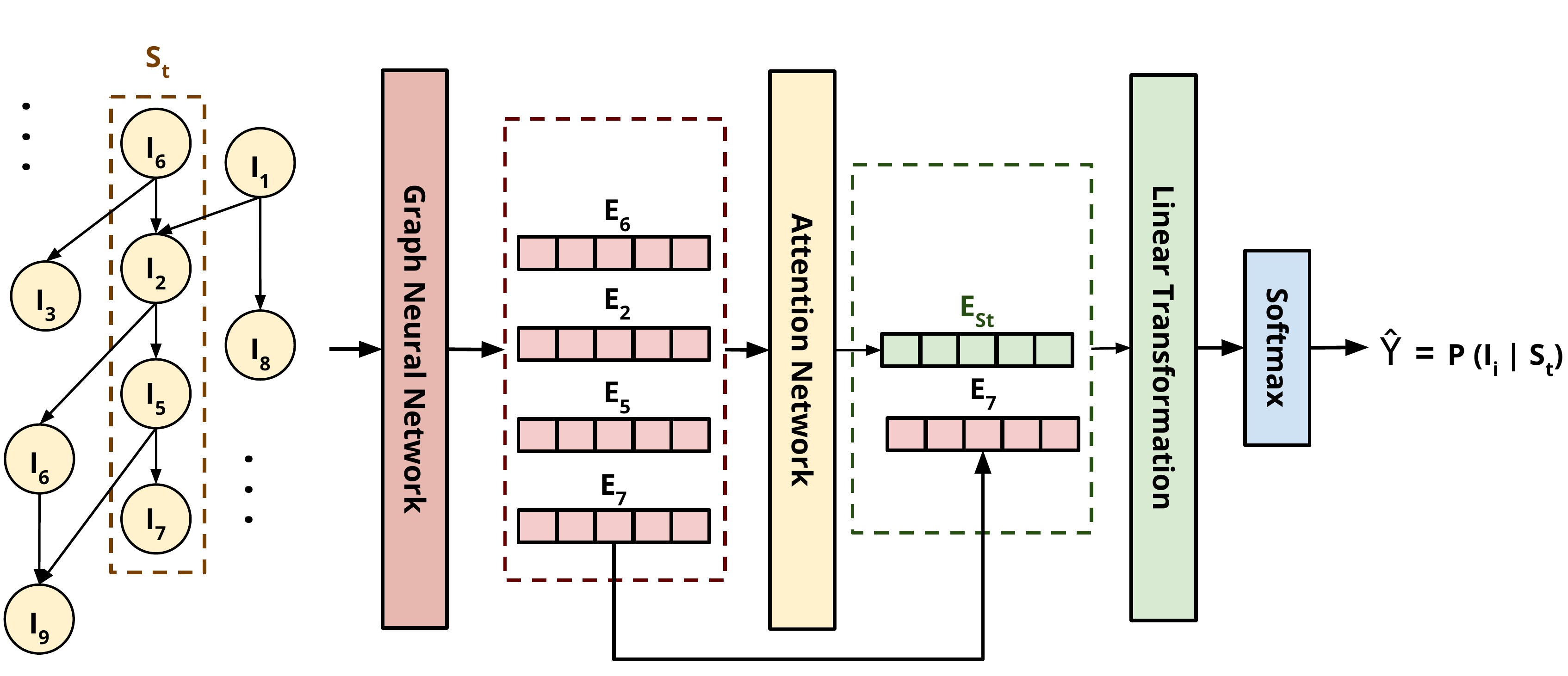}
    \caption{{SRGNN} model architecture \citep*{DBLP:journals/corr/abs-1902-07243}}
    \label{fig:SRGNN}
\end{figure}
Each session is modeled as a separate sub-graph. In this sub-graph, a node represents an item, and an edge represents a user interaction with that item. Session $S_t$ in Figure\ref{fig:SRGNN} shows as example to session sub-graph. Each edge is assigned a normalized weight calculated by the division of the edge occurrence by the out-degree of that edge's starting node. Then, using an attention network, each session sub-graph is proceeded one by one through a gated {GNN} to produce an embedding vector for each node. The role of {SRGNN} is to capture the complex transitions in the session context and generate accurate corresponding item embedding vectors. This method can be adapted if the nodes of the items have multiple features like price, color, size, and brand by concatenating them with the node embedding vector.
Further, the session embedding vector adds information about the session's local embedding vector defined by the last clicked item vector, which is $E_7$ in Figure\ref{fig:SRGNN}, and the global embedding vector $E_{S_t}$ defined by the aggregation of all the previous items vectors. This hybrid embedding approach performs a linear transformation over the concatenation of both the local and global embedding vectors, followed by a softmax layer to predict the next item probabilities.

\subsubsection{Collaborative session-based recommendation machine}
A hybrid framework applying collaborative neighborhood information to session-based recommendation was proposed by \citet*{wang2019collaborative} who hypothesized that neighborhood sessions to the current session even made by different users, can contain useful information in improving the recommendation system predictions. 
\begin{figure}
    \centering
    \includegraphics[width=0.9\textwidth]{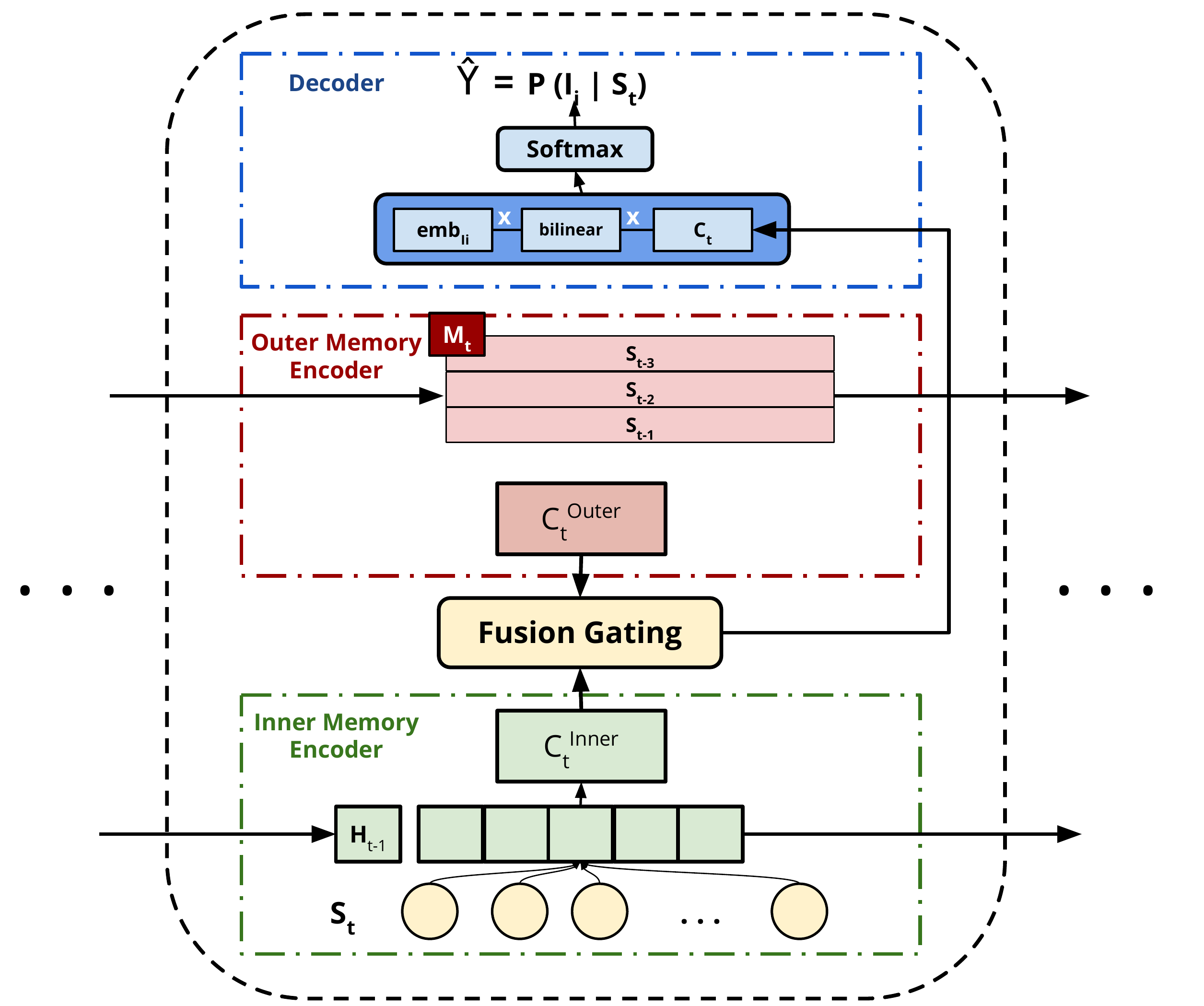}
    \caption{One unit of the {CSRM} architecture \citep*{wang2019collaborative}}
    \label{fig:CSRM}
\end{figure}

The architecture implementation, shown in Figure\ref{fig:CSRM}, includes two main encoders. First, the inner memory encoder models the user behavior during his current session using a {RNN} with an attention mechanism fed with the hidden state of the network from the previous layer $h_{t-1}$, and current session $S_t$ items. This encoder outputs two concatenated vector embeddings $C^{Inner}$ of the current session behavior representing the whole session items, and the key items clicked during the session. Second, the outer memory encoder looks for the neighborhood sessions that contain similar patterns to the current session out of a subset of the recently stored sessions $(S_{t-1}, S_{t-2}, ...)$, which are used in enhancing the recommendation process. The final output from the outer memory encoder $C^{Outer}$ represents the influence of other sessions' representations in the neighborhood memory network $M$ on the current session. The final current session representation $C_t$ is formed by a selective fusion between both encoders output. Finally, the output scores for all items are predicted using a bi-linear decoding scheme between the embedding of item $I_i$, and the final representation vector of the current session $C_t$, followed by a softmax layer.
Other two main advantages in {CSRM} are:
\begin{enumerate}
    \item Storing recent sessions and looking for neighborhoods within these sessions can be beneficial, especially in e-commerce where temporal drifts in user's interests occur frequently.
    \item Ease of including different item features in the item embedding vector, which can enhance the recommendations' accuracy.
\end{enumerate}

\section{Methodology}
\label{sec:4}

\subsection{Datasets}
All experiments were based on benchmark datasets in the e-commerce domain:
    \subsubsection{YOOCHOOSE}
    The first dataset is collected by YOOCHOOSE \footnote{https://www.yoochoose.com/} incorporation and published in RecSys Challenge 2015 \footnote{http://2015.recsyschallenge.com/}. The dataset contains a collection of sessions from a retailer, where each session includes the click events that the user performed in the session. The data was collected during $\approx 6$ months in 2014, reflecting the clicks and purchases performed by the users of an online retailer in Europe. The main characteristics that distinguish this dataset from others are having the largest number of clicks and the smallest number of items, which leads to a high presence of most of the items in the dataset. Following the previous literature, we used the last day sessions as a testing set and the rest sessions as a training set. This dataset is referred to as {RECSYS} \citep*{hidasi2015session, DBLP:journals/corr/HidasiK17}.

    \subsubsection{Diginetica}
    Diginetica dataset was used in CIKM Cup 2016 \footnote{https://cikm2016.cs.iupui.edu/cikm-cup/} for the personalized e-commerce search challenge. The dataset was provided by DIGINETICA \footnote{http://diginetica.com/} corporation containing anonymized search and browsing logs, product data, and anonymized transactions collected for five months from e-commerce websites. We used the transaction data only in our experiments. Similar to the RECSYS dataset, we used the last day sessions as a testing set and the remaining sessions as the training set. We use the name {CIKMCUP} to refer to this dataset in the rest of this paper.
    
    \subsubsection{TMall}
    TMall is a large dataset that consists of interaction logs from the e-commerce TMall website \footnote{https://www.tmall.com/}. The dataset was collected in six months, including the user-item views logs; however, the time recorded for each event was at the granularity of days. Thus, we used transactions made by the same user in one day as one session, which leads to much longer sessions than the other datasets. Due to the constraints in the computational resources, we used only the dataset in the range from the beginning of September to the end of October (two months) as the training set, and the subsequent day (1st of November) as the testing set. We refer to this dataset as {TMALL}.
    
    \subsubsection{Retail Rocket}
    Finally, the Retail-Rocket dataset was collected and published by retail-rocket e-commerce personalization company \footnote{https://retailrocket.net/} aiming to motivate researches in the field of recommendation systems. The dataset includes user behavioral data from a real-world e-commerce website throughout $\approx$ 4.5 months like views, add to carts, and transactions in addition to items identifiers and their properties in a hashed format. Only the views and add-to-cart events were considered in our experiments, while transaction events are discarded. This dataset and the CIKMCUP dataset are characterized by the small number of clicks compared to the number of existing unique items. Besides, they also have fewer sessions than both the {TMALL} and {RECSYS} datasets. We used the last two days as the testing set while the rest of the sessions as the training set.  We refer to this dataset as {ROCKET}.

During the preprocessing of all datasets, we filtered out sessions of length one as they do not include enough items for evaluation. Additionally, we filtered the clicked items in the test sets, which do not exist in the corresponding training sets in all the experiments. \emph{Multiple consecutive clicks on the same item in one session are replaced by a single click on that item}. This step was done as it does not make sense to recommend the same item currently viewed by the user, and it is always preferable to recommend new related items. For example, a session of a click sequence of (1, 1, 1, 2, 2, 3, 4, 4, 1) is pre-processed to (1, 2, 3, 4, 1). Ignoring this step, like in previous studies \citep*{ludewig2018evaluation, ludewig2019empirical}, falls in favor of the baseline methods like nearest neighbors and frequent pattern mining over neural-based methods. We kept all the items in the training set and did not remove low-frequency items. During the evaluation, we compute the accuracy of recommendations on all the possible splits starting from the first click of every single session. For instance, in a session represented by the vector (1, 2, 3, 4), we evaluate the recommendations on the session of a single click on (1) with target item 2, the (1, 2) session with target item 3, and the (1, 2, 3) session with target item 4. Finally, the average performance measurements are reported out. The statistics of the datasets after the preprocessing are summarized in Table \ref{tab:1}

\begin{table}[t]
\centering
\caption{Final statistics of the datasets used in the evaluation experiments}
\resizebox{\linewidth}{!}{\begin{tabular}{c|c|c|c|c|}
\cline{2-5}
 & {RECSYS} & {CIKMCUP} & {TMALL} & {ROCKET} \\ \hline
\multicolumn{1}{|c|}{Number of Items} & 37.48K & 122.53K & 618.77K & 134.71K \\ \hline
\multicolumn{1}{|c|}{Number of Sessions} & 7.98M & 310.33K & 1.58M & 367.59K \\ \hline
\multicolumn{1}{|c|}{Number of Clicks} & 27.68M & 1.16M & 10.83M & 1.06M \\ \hline
\multicolumn{1}{|c|}{Timespan in Days} & 175 & 152 & 62 & 138 \\ \hline
\multicolumn{1}{|l|}{Average Item Frequency} & 738.5 & 9.5 & 17.5 & 7.9 \\ \hline
\multicolumn{1}{|c|}{Average Session Length} & 3.47 & 3.75 & 6.86 & 2.88 \\ \hline
\end{tabular}}
\label{tab:1}
\end{table}

\subsection{Experiments Description}
Our study includes eight different sets of experiments repeated for each model on all the evaluated datasets. The target of these experiments is to answer the following research questions (RQs):
\begin{itemize}
    \item RQ1: Different Training Session Lengths: \\
    We aim to evaluate which models can learn from short sessions in length during the training process, and which ones can make use of lengthy sessions better to specify the user's interest accurately. To answer this RQ, we divided each training dataset into three different splits according to their length. We keep only sessions of length $<5$ in the first split, $\geq 5$ \& $<10$ in the second split, and $\geq 10$ for the third one. We choose these thresholds as it sounds challenging to determine a correct session context with $<5$ clicked items. Sessions of length $>5$ and $<10$ have an adequate number of items to determine the user's preferences. Sessions with $>10$ items have very enough items to model the user's preferences, too, but it is also more likely to have a drift in the session context that may or may not be captured by the model. This selection was also made following \citet*{liu2018stamp}, who chose a threshold of 5 to distinguish between short and long sessions as all datasets have an average session length that is close to $5$, as shown in Table \ref{tab:1}. All the models are trained on each split of the training sessions. However, the evaluation was done on the testing splits extracted from the original datasets without further pruning.
    
    \item RQ2: Different Testing Session Lengths: \\
    In this experiment, we measure the models' performance with different session lengths during inference. The main target of this experiment is to observe the model performance during the start of the session and after having an adequate number of interactions from the user. This experiment could help determine which models can not perform well at long sessions that usually include user drifts in preferences. On the contrary to the previous RQ, we fixed the same training dataset and divided the test sets to three different splits of sessions of maximum length of $5$, $10$, $>10$, respectively. All models are trained on the same training set and evaluated on each test split. 
    
    \item RQ3: Prediction of items with different popularity in the training set: \\
    In this set of experiments, we investigate how the models' performance changes concerning the items' frequency in the training set. Answering this RQ can help determine which models can learn well from less frequent items in the training set and predict them accurately during evaluation. Additionally, this experiment can show how models are biased towards predicting the more popular items. In this experiment, we divided the test sets of each dataset to keep only items whose frequencies do not exceed a specific threshold in the training set. The frequency thresholds used for different splits were ($50$, $100$, $200$, $300$, $>300$) for the RECSYS and TMALL datasets, and ($10$, $30$, $60$, $100$, $>100$) for the CIKMCUP and ROCKET datasets. This categorization was chosen based on the distribution of the items' frequency in the training sets such that each category of a range of frequencies has an adequate number of items (>1000 items) covering the whole range of frequencies, as shown in Figure \ref{fig:freq-dist}. All models are trained on the same training set, and the evaluation metrics are computed on each set of items in the testing set, satisfying the above frequency threshold conditions.
    
    \begin{figure}[!ht]
        \centering
        \includegraphics[width=\textwidth]{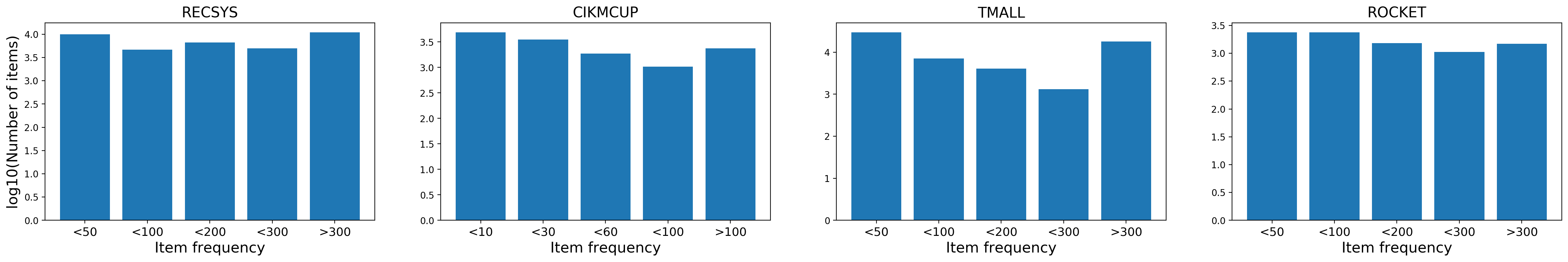}
        \caption{Frequency distribution of the items in the training sets.}
        \label{fig:freq-dist}
    \end{figure}
    
    \item RQ4: Effect of Data Recency: \\
   In this experiment, we divide our training sets into three data portions. Each portion is collected during a period that is equal in length to the others but different in terms of the creation date (recency). The first portion represents the most recent collected data sessions. The second one represents the eldest sessions, and the last one is a mixture between the more recent half of the first portion and the older half of the second portion. All models are evaluated on the same test set. We aim in this experiment to show whether it is crucial to have a dynamic time series modeling to be taken into account while fitting the different models or seasonal changes can not make a significant drift in the user preferences in the e-commerce domain. For example, in fashion e-commerce, users tend to look for light clothes during summer that makes it unlikely to learn from data collected in the winter, where users usually have more preference for heavy clothes. Additionally, it is quite essential to determine how data recency could lead to data leakage problems that could result in a deceiving model accuracy during training.
    
    \item RQ5: Effect of Training Data Size: \\
    We aim by this experiment to observe the models' performance on different dataset sizes. Answering this RQ can help understand the suitable dataset size corresponding to the number of available items in a dataset such that the model performance is not affected profoundly. Consequently, saving more computational resources without impairment in the models' performance.  We have selected randomly different splits of the original training datasets such that the sizes of these splits are equal to $\frac{1}{P}$ of the original training set size where $P \in {2,8,16,64,256}$.
    
    \item RQ6: Effect of Training Data Time Span: \\
    Additionally, to check whether the results obtained from RQ4 and RQ5 can be generalized or not, we run a similar experiment to RQ5; however, instead of selecting random portions of training sets, we divided them according to the time taken in the data collection. For example, we used only the most recent sessions collected during the last $m$ days before the period used as a testing set to train the model. In this experiment, we used $m \in {2, 7, 14, 30}$, and we aim to know the time span required to train different models and achieve the best performance according to the different data set properties like the number of items, and average session length.
    
    \item RQ7: Items Popularity and Coverage: \\
    What are the coverage and popularity of the items of each model on fixed dataset splits? Given the models' predictions, we compute the coverage and popularity of these predictions out of the total number of unique items in the dataset. These measurements can provide a good indication of the models' tendency to predict the most frequent items only, or they can cover the space of items to a large extent. The coverage of the model predictions is a measure of what is called aggregate diversity and how the model is adapted to different sessions' context \citep*{shani2011evaluating}. A small coverage value shows that the model is always recommending a small set of items for all users, as the most popular or frequent items in the training set. A high coverage shows that it recommends a wide range of items with different sessions context \citep*{ludewig2019empirical}. The coverage can also be shown and confirmed by the popularity metric that computes the average frequency of occurrence of the predicted items from the training set normalized by the frequency of the most popular item. We used the full original training and testing sets of {CIKMCUP} and {ROCKET} datasets. Besides, we used a random split of $\frac{1}{16}$ of the {RECSYS} and {TMALL} datasets for training with the last 2-days sessions for testing. The coverage and popularity of the top five predicted items per session are reported in each of these experiments. Models with high accuracy in terms of HR and MRR and high coverage of items are usually preferred over accurate models with lower coverage. This reflects how the model provides different predictions that are adaptable to the context of the sessions.
    
    \item RQ8: Computational Resources: \\
    What are the required time and memory resources for each model during both training and inference phases? In this experiment, we report the different computational resources required by each model to observe the trade-off between the model performance and its complexity and if it is worth having more computationally expensive models than simpler ones. Besides, we aim to find the suitability of different models to be used practically in making real-time predictions.
\end{itemize}

The properties of all training and testing splits used in each experiment are summarized in Table \ref{tab:appendixtab0}. We used an early stopping approach during training with a validation split of 10\% of the training split for the deep-learning models. Additionally, as hyperparameter optimization is an essential part in determining the performance of models, we run a random search of 20 iterations for all models on each dataset to tune the most effective hyper-parameters suggested to be tuned by their authors or based on our own experiments. However, we kept the rest of the networks' hyper-parameters by their default values, as mentioned in their corresponding papers. We selected the hyper-parameter settings achieving the highest $HR@20$ for each dataset. The list of the tuned hyper-parameters for each model, along with their ranges, can be found in Table \ref{tab:appendixtab19}.

Our work was carried out in part in the high performance computing center of the University of Tartu\footnote{https://hpc.ut.ee/} In case that graphical processing unit (GPU) is used for neural network models, we used NVIDIA Tesla P100 GPU. The memory size was limited to 20GB RAM, Intel(R) Xeon(R) CPU E5-2660 v2 @ 2.20GHz processors with up to 30 cores were allocated from the computing center to run the models that do not support GPU. During reporting training and testing time and memory consumption in our results, we have not used any GPUs to make the comparison fair among all models. However, indeed all neural-based models support the usage of GPUs, which is a significant advantage over other algorithms. The source codes used in this study, and the logs of the results, are made publicly available \footnote{https://github.com/mmaher22/iCV-SBR}.

\subsection{Evaluation Metrics of models performance}
We measured the performance of all models in our experiments using several evaluation metrics:
\begin{enumerate}
    \item Hit Rate(HR@K) is the rate of matching the correct item clicked by the user with any of the list of predictions made by the model. The metric value is set to 1 if the target item is present among top K predictions and 0 otherwise. The formula of $HR@K$ for a dataset $D$ is described as follows: 
    \begin{equation}
    \label{eq:hr}
        HR@K = \frac{1}{|\mathcal{S}_{D}|} \sum_{i=1}^{|\mathcal{S}_{D}|} 1_{IN}\left(I_{target}, \argmax_{K} (\hat{\mathbf{Y_i}}) \right).
    \end{equation}
    \item Mean Reciprocal Rank(MRR@K) is the average of reciprocal ranks of the target item if the score of this item is among the top K predictions made by the model. Otherwise, the reciprocal rank is set to zero \citep*{hidasi2015session}. The computation of $MRR@K$ is given by
    \begin{equation}
    \label{eq:mrr}
        MRR@K = \frac{1}{|\mathcal{S}_{D}|} \sum_{i=1}^{|\mathcal{S}_{D}|} \frac{1}{r_{IN} \left(I_{target}, \argmax_{K}(\hat{\mathbf{Y_i}}) \right)}.
    \end{equation}
    \item Item Coverage(COV@K) is the measurement of how the model predicts a variety of items and not only biased to a small subset of frequent items. The item coverage is the ratio of the number of unique items predicted by the model to the total number of items in the training set. Given that $K$ is the number of top predictions to be considered from the model for each session, item coverage is described as
    \begin{equation}
    \label{eq:cov}
        \text{COV@K} =  \frac{\left| \mathcal{U} \left( \argmax_{K} \left(\hat{\textbf{Y}}_{i=1,2,...,|\mathcal{S}_{TE}|} \right) \right) \right|}{|\mathcal{U}(\mathcal{S}_{TR})|}.
    \end{equation}
    \item Item Popularity(POP@K) is a representation of how the model tends to predict popular items. This metric can reveal models that achieve good performance based on the popularity of certain items in the training set instead of recommending items that match the session context and the user's preferences. The item popularity is the ratio between the average of the predicted items' frequencies to the frequency of the most popular item in the training set. Given that $K$ is the number of top predictions to be considered from the model for each session. Item popularity is given by
    \begin{equation}
    \label{eq:pop}
        POP@K = \frac{\sum^{|S_{TE}|}_{i=1} \left( \sum_{I \in \argmax_{K}(\hat{\mathbf{Y}}_{i})}{\mathcal{F}}(I) \right)}{|S_{TE}| \cdot {max \left(\mathcal{F}(S_{TR}) \right)}}. 
    \end{equation}
\end{enumerate}

\section{Results}
\label{sec:5}
In this section, we report and discuss the results obtained from our extensive evaluation for the different models trying to answer the different research questions proposed in Section \ref{sec:4}

\begin{figure}[!ht]
    \centering
    \includegraphics[width=\textwidth]{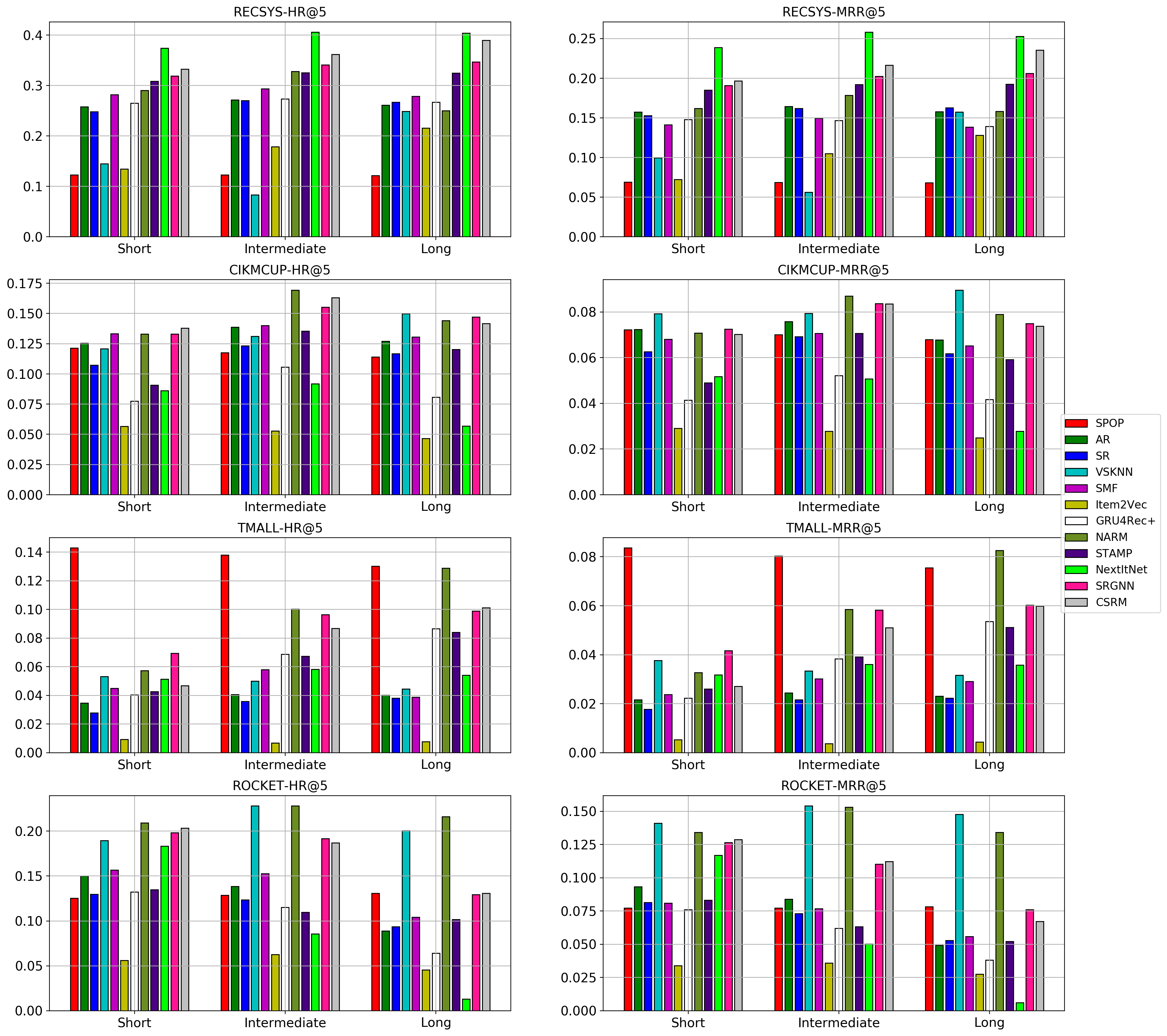}
    \caption{RQ1: Effect of using different training session length on algorithms performance.}
    \label{fig:Train_Length}
\end{figure}

\subsection{RQ1: Different Training Session Lengths:}
In the presented diagrams, we report the results for HR and MRR of each model on all the evaluated datasets. On the contrary to some previous literature work like \citet*{ludewig2018evaluation}, who used predictions cut-off threshold of 20 recommendations, We have chosen to set the number of predictions cut-off to 5 as it is more reasonable to recommend around five items to the user in a real use case. Also, 20 recommendation is a large number to be used in real-life e-commerce scenarios. However, full results for different predictions cut-off thresholds (1,3,5,10,20), in this experiment and all the following ones, can be found in our online repository \footnote{\label{repo}https://github.com/mmaher22/iCV-SBR/tree/master/Results}.

As shown in Figure \ref{fig:Train_Length}, most of the neural models outperform the non-neural baseline models except for {S-POP}, which is the best model in the {TMALL} dataset due to the nature of the dataset where the clicked items in one session are repeated more frequently within the same session than other datasets. TMALL has an item frequency per session of $1.204$ on average compared with $1.097, 1.115, 1.103$ for RECSYS, CIKMCUP, and ROCKET, respectively. This difference means that it is more likely that the same item appears multiple times in a single session in the TMALL dataset. Also, VSKNN has a relatively good performance in the ROCKET dataset. However, the top three models in terms of either HR or MRR in RECSYS and CIKMCUP datasets are always neural-based. Besides, two neural models are always among the top three performing models in the TMALL and ROCKET datasets. {NextItNet} has the highest performance in the {RECSYS} dataset characterized by the highest average item frequency among the used datasets. This property is suitable more for convolutional networks that require a large number of sessions covering all items to model them correctly. However, there is a small decrease in the performance of {CSRM}, {SRGNN}, and {STAMP} when training using short session length, which is apparent in {RECSYS} and {TMALL} that have mostly intermediate to long sessions in the corresponding testing sets.
On the other hand, {GRU4Rec+} has the highest performance when training using intermediate and short session length. In contrast, this performance degrades on long sessions since a drift in the user preferences is more likely to occur. Although {NARM} also uses a similar network to {GRU4Rec+}, it has a better performance thanks to the attention layers in its architecture. This improvement in performance was significant compared with a baseline of the GRU4Rec network, that suffers from the vanishing gradient problem, especially in long sessions of 11 to 17 clicks \citep*{li2017neural}.

Regarding baseline models like {S-POP}, {AR}, and {SR}, there is a big and consistent change in their performance while changing the training sessions' length, especially in the {RECSYS} dataset. However, it is clear that these models have comparable performance to neural-based models in datasets where the average item frequency is small and not enough for learning good item representations like in {CIKMCUP}, {ROCKET} and {TMALL} datasets. Overall, baseline methods like S-POP, VSKNN, and SMF are more suitable for short sessions when there are not enough session events to represent the user's preferences. On the other hand, as the session includes a larger number of events, more complex models become better than baselines. This improvement is only apparent for intermediate sessions compared with short sessions. However, long sessions have similar or slightly worse performance since the user preferences are more likely to change in the long sessions. The number of these sessions is not large enough to train the neural models well, especially in the CIKMCUP and ROCKET datasets. However, the total number of events in all training splits of these datasets is close since long sessions have higher average session length than short ones, as shown in Table \ref{tab:appendixtab0}. In contrast, the TMALL dataset has many more events in the long sessions than intermediate and short ones. This difference in splits' size could contribute to the noticeable improvement of the performance of the neural models trained with long sessions like NARM, as also discussed in Section \ref{sec:size-effect}.

\begin{figure}[!ht]
    \centering
    \includegraphics[width=\textwidth]{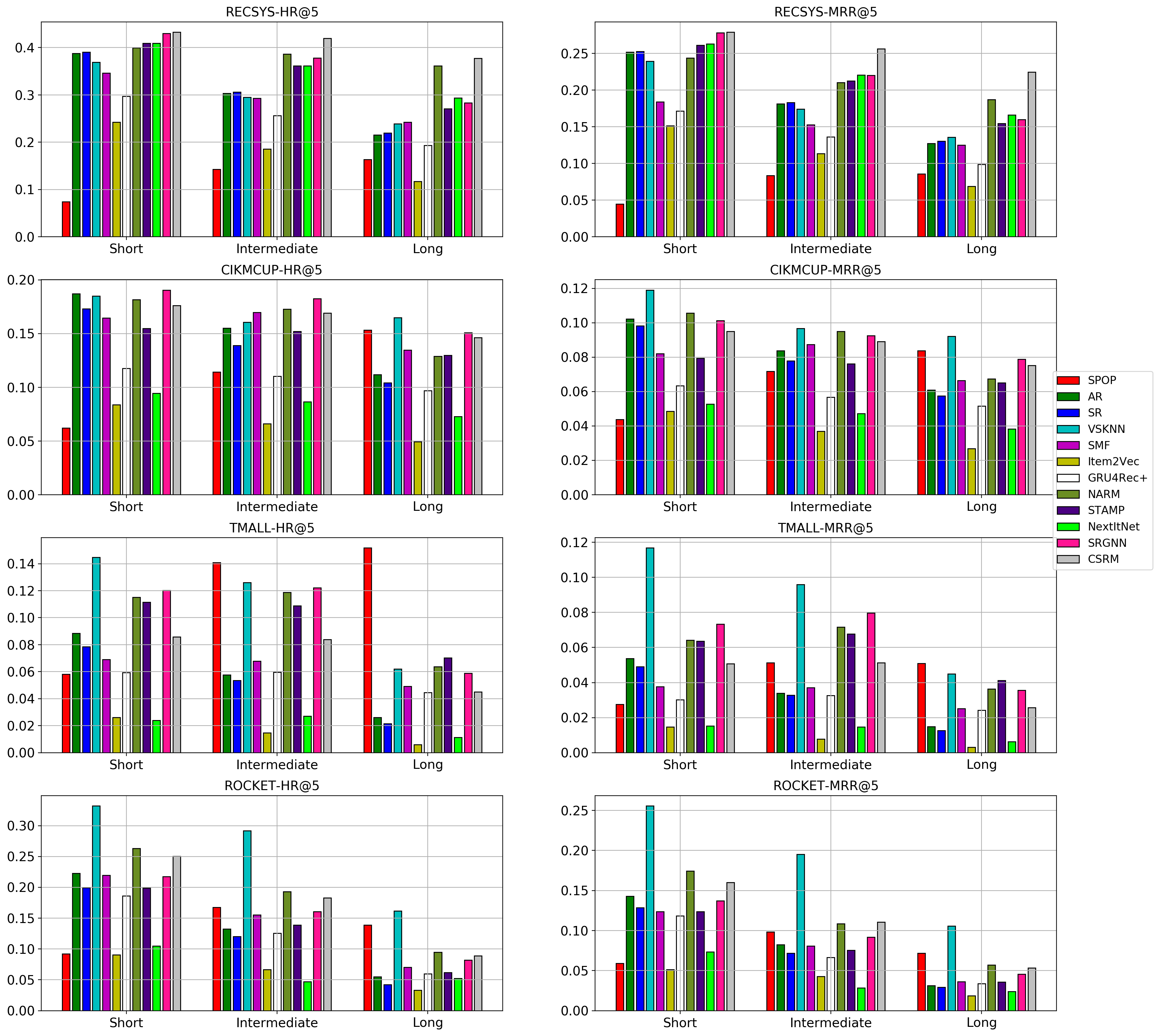}
    \caption{RQ2: Performance of the models on different testing session lengths.}
    \label{fig:Test_Length}
\end{figure}

\subsection{RQ2: Different Testing Session Lengths:}
While training using different session lengths gives useful insights about the performance of models, this performance is still highly correlated with the length of the testing sessions. For example, If a model is trained with short sessions while most of the testing set sessions are long, it will not make accurate predictions.

Figure \ref{fig:Test_Length} shows the performance of different models using the same training set for all of them while choosing a subset of sessions according to their length as a testing set. {S-POP} has an increasing performance while the session goes longer since longer sessions have a higher probability that the user clicks again on a previous item that he liked before during the same session. However, in lengthy sessions, personalization still pays off since the session has adequate information to model the user preferences precisely \citep*{quadrana2018sequence}. 

The performance of {SR} and {AR} degrade consistently by a large degree, for all the datasets as the session length increases. This impairment is due to the small window of interest that both these algorithms look at while computing the frequent patterns of items, which means that they will not make use of longer sessions. Although the window size for the computation of frequent patterns could be increased, the computational complexity grows exponentially, making it infeasible to use them with long windows sizes. The same effect holds for VSKNN when the selected weighting function for the clicked items in the session gives much higher weights to the recent items like the quadratic, multiplicative inverse, and log weighting functions. Thus, the weights given to the remaining items outside a specific window size are almost neglected. Besides, almost all the neural models' accuracy decreases at long sessions, which shows that it is still one of the challenges to model the user drift of interests during the same session. This decrease in performance in very clear in NextItNet, {Item2Vec}, {GRU4Rec+} among all datasets while it is slightly less observable for {CSRM}, {STAMP} and {NARM} which could be due to the memory and attention mechanisms applied in these models. NextItNet never outperforms in the RECSYS dataset compared with the results obtained in RQ1. This decrease in performance is due to the smaller training split size used in the experiments of RQ2 compared with the ones used in RQ1. This claim is also confirmed in the experiments of RQ5 and RQ6, where the deep architecture of convolutional layers in NextItNet requires many instances per item to model it adequately compared with attention networks.

Overall, it seems that achieving excellent performance in long sessions is still a challenging problem for most of the models. On the one hand, pattern mining models do not pay attention to long sessions. They only make use of a small window frame around the item of interest while ignoring information made earlier by the user in the same session. On the other hand, neural models can not still detect user-drifts accurately and might suffer from vanishing gradient problems in {RNN}s, especially for the very long sessions \citep*{hidasi2018recurrent, li2017neural}. Interestingly, much research has focused on a solution to the cold-start problem when users do not make enough clicks to capture their preferences. However, it seems that more attention needs to be also paid to improve the recommendation performance for the long sessions.

\begin{figure}[!ht]
    \centering
    \subfloat[$HR@5$]{{\includegraphics[width=0.48\textwidth]{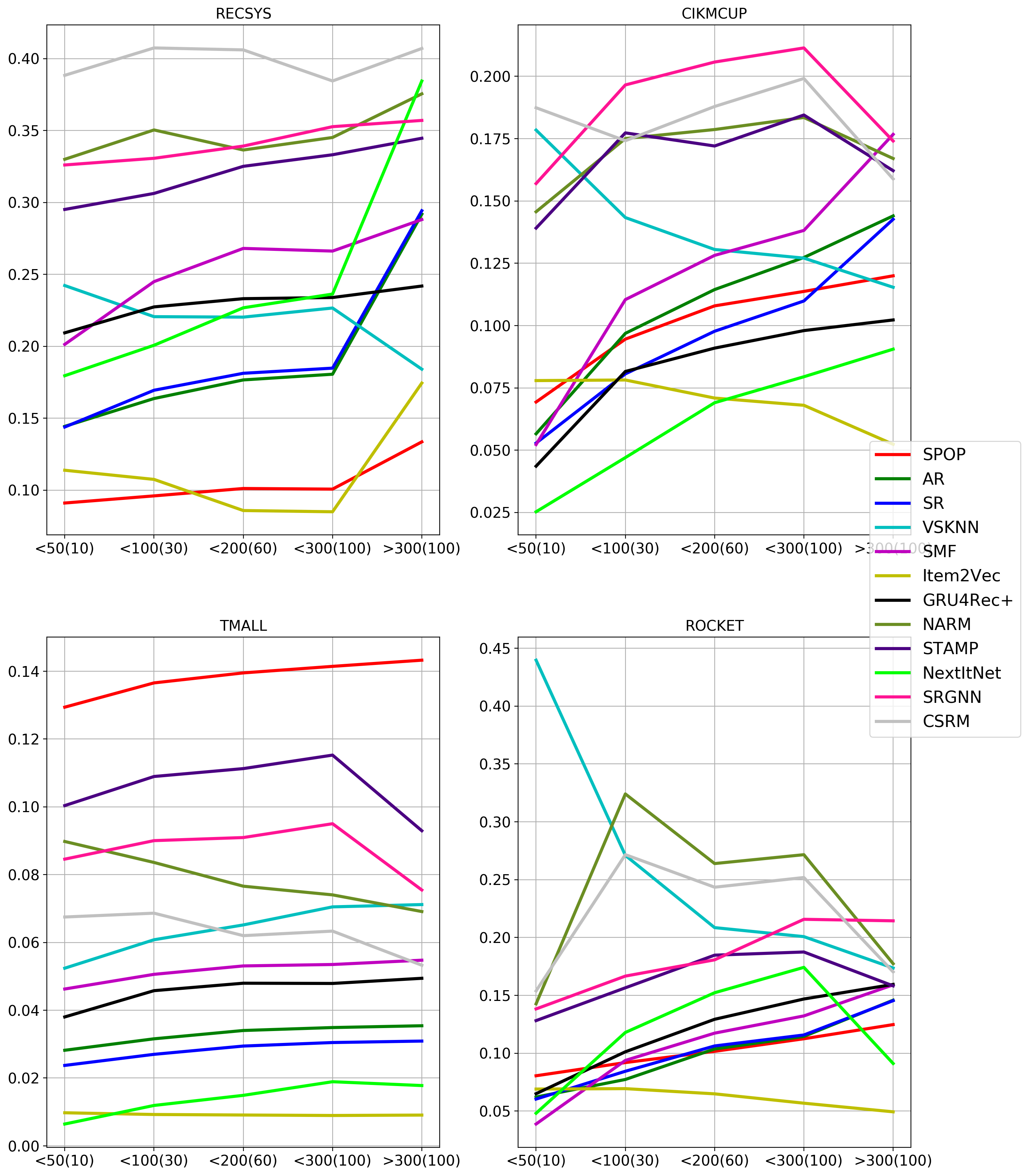} }}
    \qquad
    \subfloat[$MRR@5$]{{\includegraphics[width=0.45\textwidth]{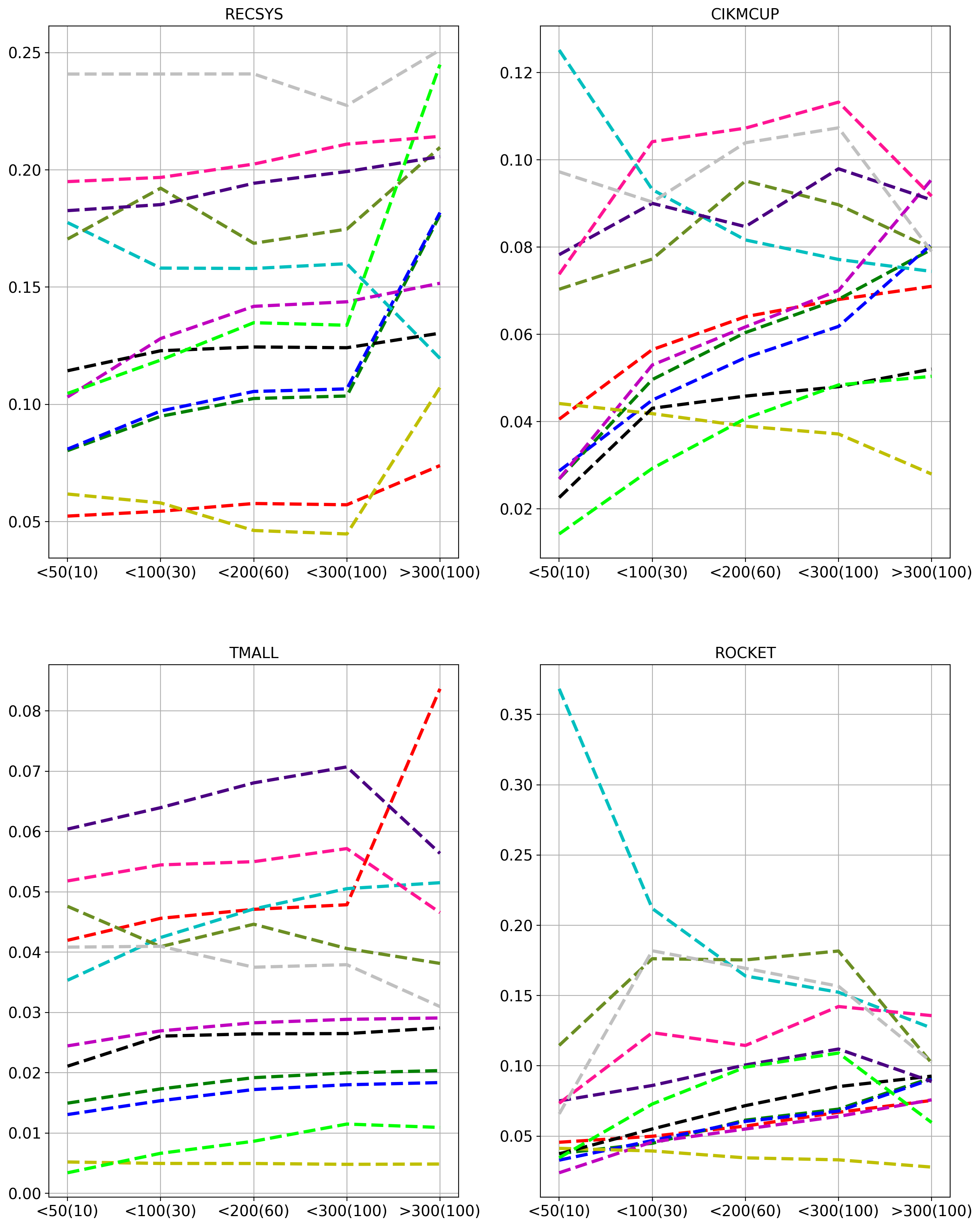} }}
    \caption{RQ3: Performance of different models using different items frequency values in training set ($<50,<100,<200,<300,>300$) for RECSYS and TMALL and ($<10,<30,<60,<100,>100$) for CIKMCUP and ROCKET.}
    \label{fig:freq}
\end{figure}

\subsection{RQ3: Prediction of items with different abundance in the training set:}
Figure \ref{fig:freq} shows the performance of different models when trying to predict an item below a specific threshold of occurrences in the training set. This experiment shows how model performance is affected by the number of items' occurrences during fitting. In this experiment, we used frequency thresholds of (\textless $50$, \textless $100$, \textless $200$, \textless $300$) for {RECSYS} and {TMALL} datasets which have higher average items frequency, and (\textless $10$, \textless $30$, \textless $60$, \textless $100$) for {CIKMCUP} and {ROCKET} datasets.

{AR}, {SR}, {SMF}, {GRU4Rec+} performance is always improved by increasing the frequency threshold among all datasets. To a less extent, {NextItNet} and {SRGNN} has a slight gain in performance by increasing frequency threshold where this gain is stopped at very high frequencies like in {TMALL} and {ROCKET} datasets. On the other hand, {NARM}, {STAMP}, {CSRM}, {Item2Vec}, and {VSKNN} performance measurements do not have a consistent trend while increasing the frequency threshold. In general, some models can have a better performance by increasing items' occurrences in training set to be able to model these items accurately like {AR} and {SR}. On the contrary, some other models do not need this high frequency of occurrences, and it is enough to be represented only a few times in the training set like {NARM}, {STAMP}, and {CSRM} which are all having various attention mechanisms.

\begin{figure}[!ht]
    \centering
    \includegraphics[width=\textwidth]{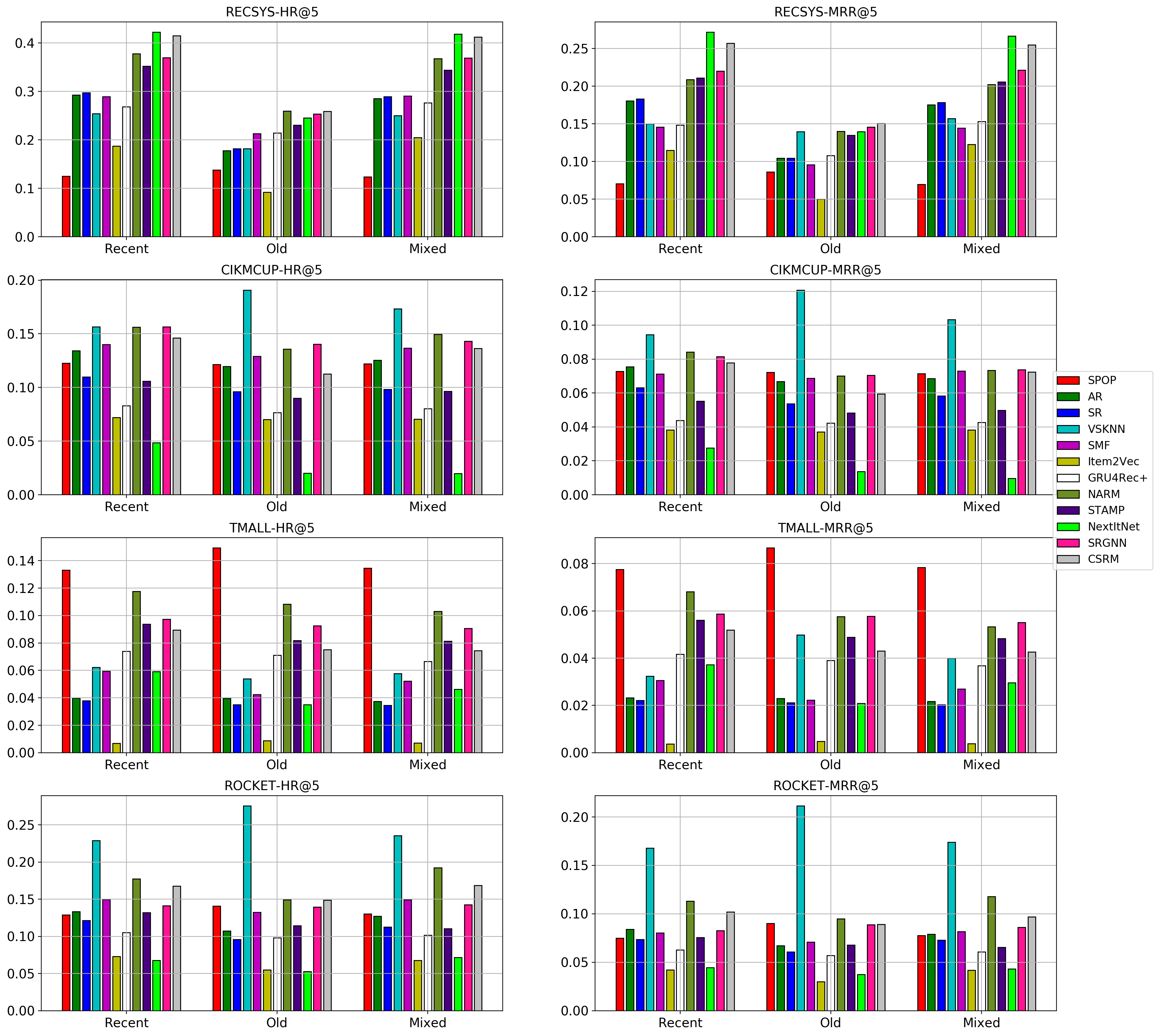}
    \caption{RQ4: Effect of data recency on models performance for different datasets.}
    \label{fig:Recency}
\end{figure}

\subsection{RQ4: Effect of Data Recency:}
Using session-based recommendation models in e-commerce always requires being up-to-date with enough recent data to model the current users' trends. In this experiment, we tested our hypothesis by training the models using the sessions collected from the most recent five days, the eldest five days, and a mix between half of recent and half of the old splits \footnote{In {ROCKET}, we used ten days instead of five as the dataset is smaller than the rest}. The test set was fixed for all these different training splits. In Figure \ref{fig:Recency}, it is shown that it is always preferable to train the models using the most recent sessions. It is consistent among all datasets that old sessions have an observable lower performance along almost all the models than recent and mixed splits. Although, there is no large difference between the models' performance on the recent and mixed splits especially in {RECSYS} and {ROCKET} datasets, there is still a small difference in favor of recent splits for {CSRM}, {SRGNN}, and {NARM} in {CIKMCUP} and {TMALL} datasets. Surprisingly, {VSKNN} is the only model with higher performance on old splits in two out of the four datasets and comparable performance in the other two datasets. This behavior could be interpreted as the algorithm only cares about neighboring sessions of exactly similar items as those clicked by users, which indeed results in better recommendations if matching sessions were found. Besides, VSKNN has a better overall performance than other models in the ROCKET, and CIKMCUP datasets as both of them are characterized by a lower average item frequency than the RECSYS and TMALL datasets as discussed in RQ3. It is worth mentioning that although a similar trend is observed for the models over each dataset, the differences in each dataset are not equal because some datasets span different time periods. For example, GRU4Rec+ has a higher HR@5 performance in the recent split than the old split by $\sim 7\%$ in the RECSYS dataset that spans six months. However, this difference is just $\sim 0.5\%$ in the TMALL dataset that spans two months. Hence, the time difference between old and recent splits in RECSYS is much bigger than that of the TMALL dataset, which could explain the differences in performance among all datasets.

These results confirm that we should account for time-series dynamic modeling in the session-based recommendation to model the trends in users' preferences. Besides, in case that the collected data is much old, there is a high chance that the nearest neighbor algorithms outperform other models.

\begin{figure}[!ht]
    \centering
    \includegraphics[width=\textwidth]{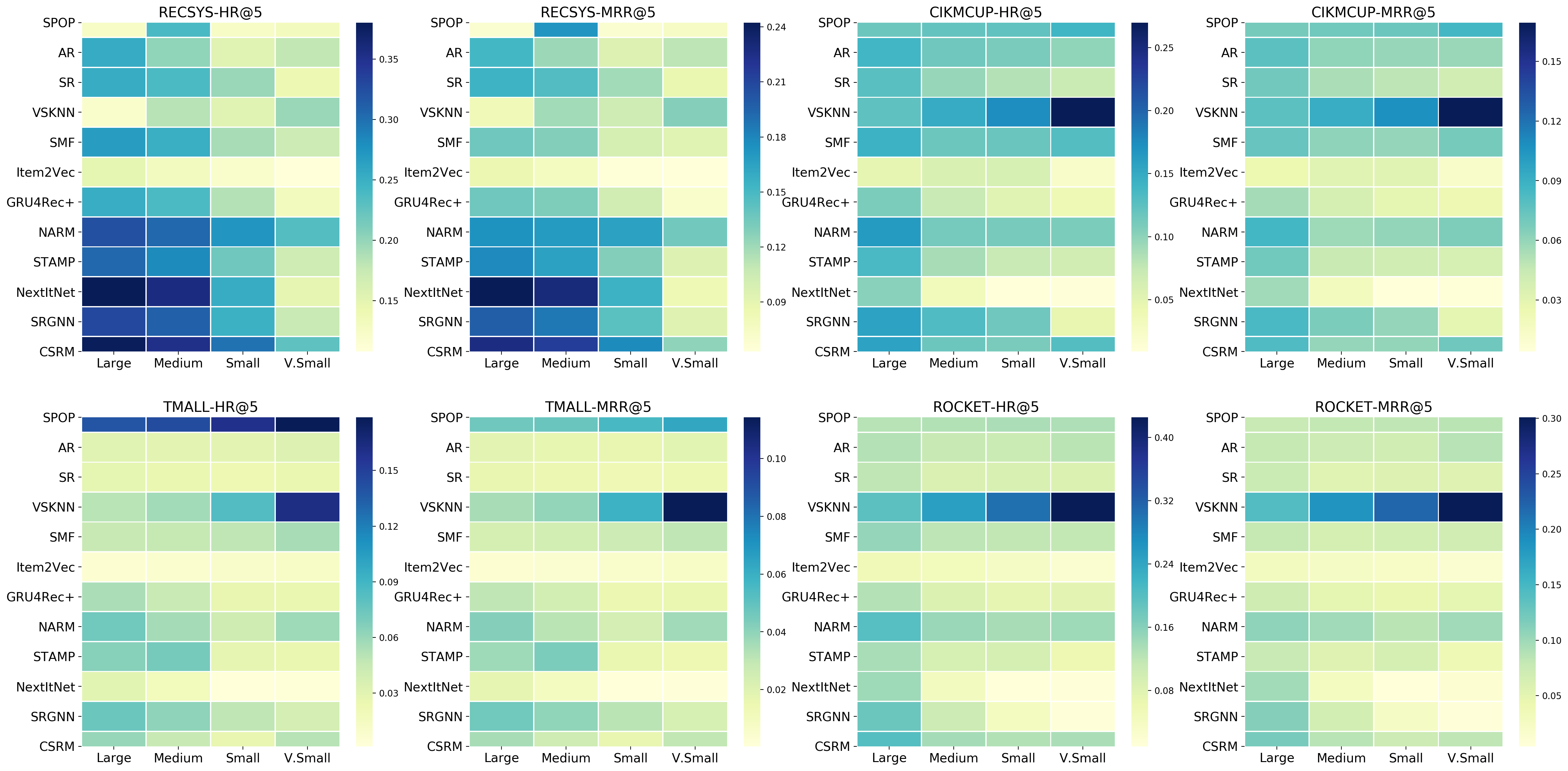}
    \caption{RQ5: Heat map of the models' performance on different training set sizes (darker means better performance). The value of each color code can be mapped to the corresponding numerical value from the vertical bar beside each subplot.}
    \label{fig:Data_Size}
\end{figure}

\subsection{RQ5: Effect of Training Data Size:}
\label{sec:size-effect}
In this experiment, we aim to know what are the suitable training data sizes corresponding to the different datasets with various characteristics. Additionally, investigating how the evaluated models perform while using these different training sets sizes. We divided {ROCKET} and {CIKMCUP} into splits of ($\frac{1}{2}$, $\frac{1}{8}$, $\frac{1}{16}$, $\frac{1}{64}$) of original training set size. {TMALL}, and {RECSYS} were divided into ($\frac{1}{8}$, $\frac{1}{16}$, $\frac{1}{64}$, $\frac{1}{256}$) splits as they are bigger. We refer to these splits as (large, medium, small, very small) respectively. 

Figure \ref{fig:Data_Size} shows a heat map for the $HR@5$ and $MRR@5$ of different models while increasing the training set size. {S-POP} and {VSKNN} are the only algorithms that do not get benefit from larger data sizes. It can be easily observed that {VSKNN} achieved its highest performance along with the four datasets when using the very-small training data portions while {S-POP} has the same behavior except for {RECSYS} dataset. All the neural models' performance is increased when using more training data sessions, which agrees with the nature of deep learning models that are data-hungry. However, {SRGNN}, {NextItNet}, and {GRU4Rec+} are consistently getting better when increasing training data sizes over all the datasets, while {NARM}, {STAMP}, and {CSRM} are improved less than the former models. Although there is a small improvement in the performance of {AR}, {SR}, and {SMF} in {RECSYS} dataset, this improvement is not clear enough in other datasets to generalize the same observation.

\begin{figure}[!ht]
    \includegraphics[width=\textwidth]{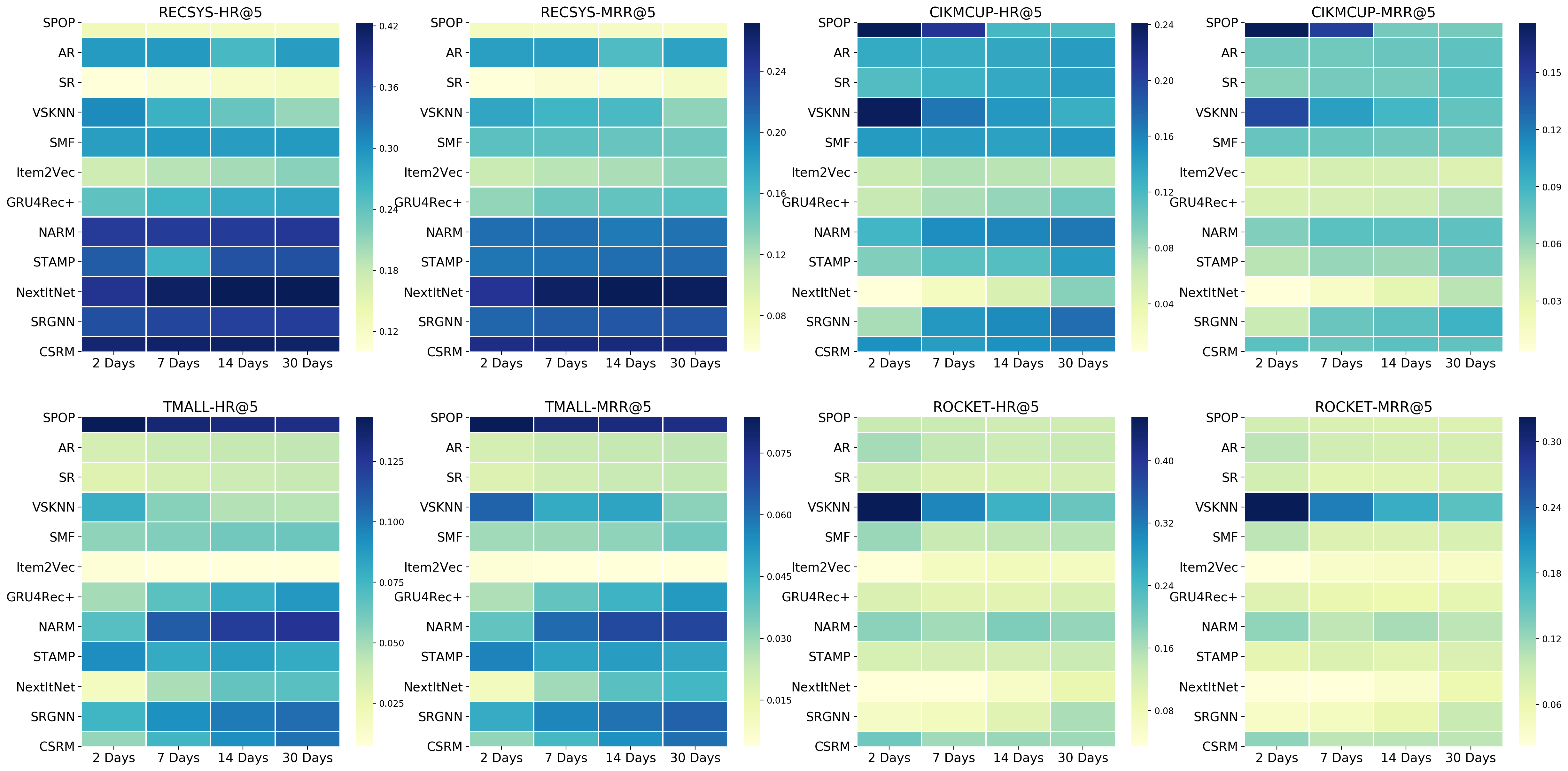}
    \caption{RQ6: Heat map of the models' performance on different training set time spans (darker means better performance).}
    \label{fig:Time_Span}
\end{figure}

\subsection{RQ6: Effect of Training Data Time-span:}
Getting some insights out of RQ4 and RQ5 about the importance of training data recency and sizes should reveal enough information about the length of the time span required to collect training data sessions. Similar to  RQ5, Figure \ref{fig:Time_Span} illustrates a heat-map of the $HR@5$ and $MRR@5$ metrics when training using splits of the most recent $x$ days from the full training set for each dataset, where $x = 2, 7, 14, 30$. Similar to what is shown previously in RQ5, in Figure \ref{fig:Data_Size}, {VSKNN}, and {S-POP} still have the best performance when training using a time-span of just two days. Additionally, the performance of neural models becomes better when increasing dataset time-span. However, in {RECSYS} dataset, the model improvement is almost ceased as it has a small number of items, and the number of sessions in a 2-days time-span is quite enough to to be used in modeling the context of the sessions accurately.

\begin{figure}[!ht]
    \centering
    \includegraphics[width=\textwidth]{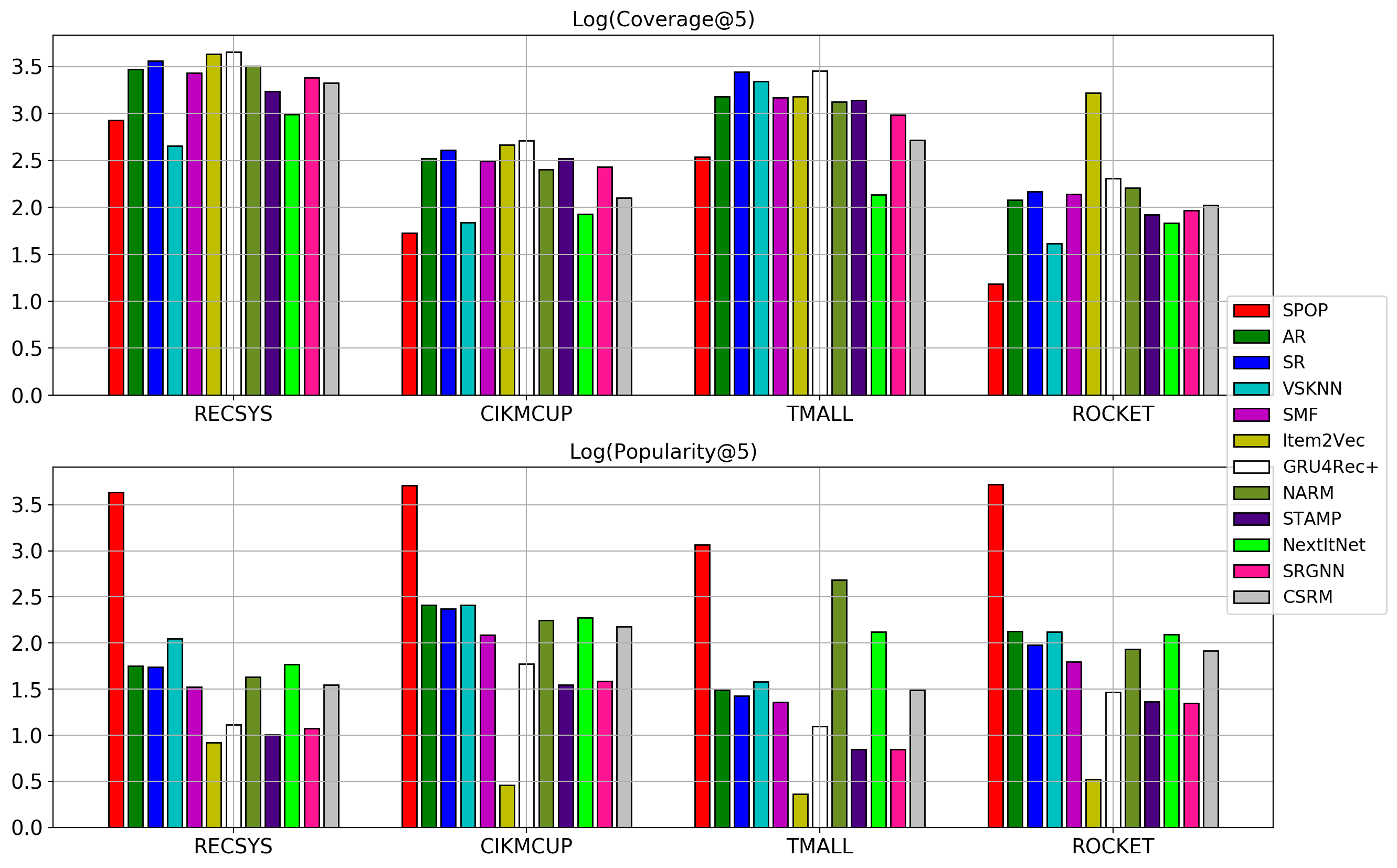}
    \caption{RQ7: Items Coverage and Popularity of Models Predictions}
    \label{fig:Cov_Pop}
\end{figure}

\subsection{RQ7: Items Coverage and Popularity:}
Item coverage and popularity are good indications of how models tend to cover the space of items in training set in making recommendations. A model with small coverage and high popularity means that it tends to predict the same items for all users, regardless of the session's context. Figure \ref{fig:Cov_Pop} shows the natural logarithm of items coverage ($COV@5$) and popularity ($POP@5$) using the same training and testing splits for each of the datasets. In general, a similar trend is observed among all the datasets comparing the baselines and neural models. For instance, {S-POP} has the lowest coverage and highest popularity since it predicts only the most frequent items. On the other hand, {Item2Vec} has the highest coverage and lowest popularity. However, this is not the case in most real-life scenarios. There are usually some popular items that the users usually click on, like the items with high discounts. So, Item2Vec still has the lowest performance in terms of HR and MRR since its output vectors are usually dispersed in the vector space, and simple distance measurements are not enough to capture the similarity among the session context and vectors of the items \citep*{kim2017bag}. Regarding baselines, {AR}, {SR}, and {VSKNN} have quite similar coverage and popularity except for {CIKMCUP} where {VSKNN} has higher item coverage. {SMF} has quite smaller both item coverage and popularity.

Regarding the neural models, {GRU4Rec+} has slightly higher coverage for its predictions, followed by {NARM}, {STAMP}, {SRGNN}, and {CSRM} but these differences are too small and can be barely observed. CSRM has a memory for storing the most recent sessions, and it predicts items based on the neighborhoods within these sessions. So, it is always biased towards a subset of recently clicked and popular items than other models. On the other hand, {NextItNet} has the lowest item coverage with comparable popularity to {NARM} and {CSRM}, which suggests that this model is more likely to get over-fitted to a small subset of items, and needs better regularization approaches to be applied to the model. NextItNet is characterized by the presence of the convolutional filters in its architecture, which require much more occurrences per item to generalize well compared with the attention-based networks \citep*{barry2018evaluation}. {SRGNN} and {STAMP} have quite smaller average popularity across their predictions than other neural models over all the datasets. Besides, NARM has a high item coverage and high popularity with a relatively high accuracy performance according to the HR and MRR metrics. This performance suggests that NARM has the advantage of recommending a wide range of items according to the different sessions' context. However, using SRGNN and STAMP could still be preferable if they have similar accuracy performance with other models as they cover more unpopular items in their recommendations. Detailed results for other predictions thresholds in this experiment can be found in Table \ref{tab:all} in the appendix. 

\begin{figure}[!ht]
    \centering
    \includegraphics[width=\textwidth]{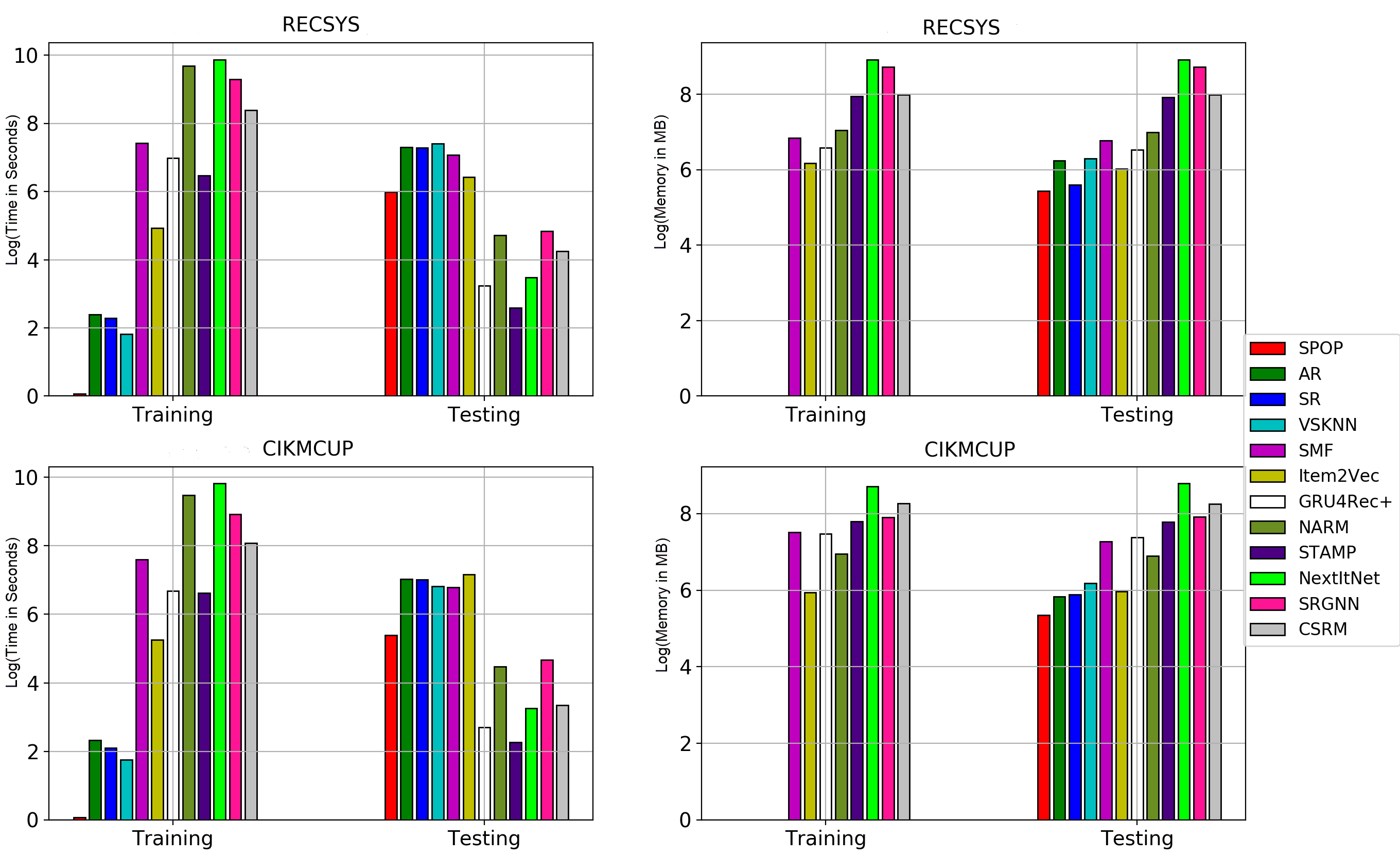}
    \caption{RQ8: Models time and memory consumption on RECSYS and CIKMCUP datasets during training and testing.}
    \label{fig:Resources}
\end{figure}

\subsection{RQ8: Computational Resources:}
It is quite important to have a short testing time to make predictions quickly as it is required to provide the user with recommendations in real time after making a specific action. Simultaneously, training computational complexity is important in terms of the scalability of the model and the ability to train it easily every short period of time. Figure\ref{fig:Resources} summarizes the computational complexity of different models during both training and testing phases in {RECSYS}, and {CIKMCUP} datasets. {S-POP}, {AR}, {SR}, and {VSKNN} are instance-based algorithms where the learning process occurs during inference by iterating over the training set for each test instance. Consequently, the computational resources for training these models are almost neglected. On the other hand, they take a very long during inference, which means that they are not suitable to be employed in making real-time predictions. However, they do not consume much memory as only the dataset, and very few parameters are required to be stored. {Item2Vec} and {SMF} have quite long training and testing time. Although Item2Vec is considered as a neural model, during inference, the similarity distance is computed between the predicted session embedding vector and the items' vectors, which takes much time. Besides, SMF is a matrix factorization algorithm that performs heavy matrix multiplication operations during both training and inference. These operations are computationally expensive in terms of both time and memory consumption.

Regarding neural models, all of them have relatively high training time and memory resources; however, they are still characterized by a short time during inference. they only need only a single forward pass to make predictions for one batch of instances. This performance suggests that neural-based models are suitable for real-time predictions. The differences in the training and inference time of the neural models are proportional to the size of each network, the number of layers, and the types of these layers. {STAMP} has the lowest training and testing time consumption as it is the smallest model in size, followed by {GRU4Rec+}, {CSRM}, then {NARM}, {SRGNN} in ascending order. {NextItNet} that is characterized by the presence of multiple convolutional layers and relatively large model has high memory consumption due to the mapping of the sessions into images. However, {NextItNet} has a smaller testing time compared with {NARM}, {SRGNN}, and {CSRM} due to the weight sharing properties of the convolutional layers. Additionally, SRGNN, a graph neural network, has a relatively large memory and time consumption due to the large size of the graph network created by mapping the items and sessions into the corresponding nodes and edges. Overall, all neural models are more compatible with the requirements of real-time predictions. However, they need ample computational resources during training using the back-propagation scheme compared with the simple baseline algorithms.

\begin{figure}[!ht]
    \tiny
    \centering
    \includegraphics[width=\textwidth]{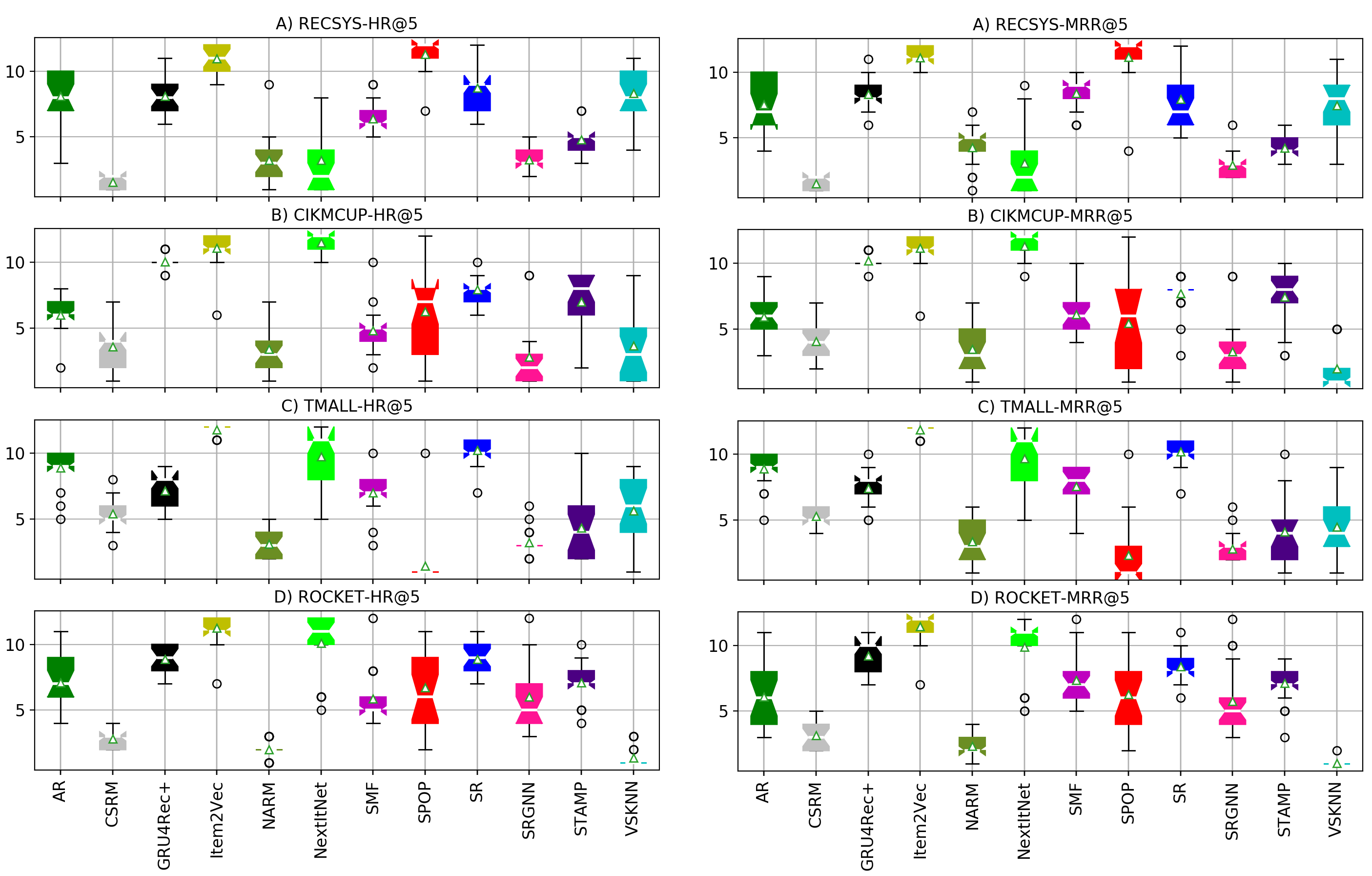}
    \caption{Ranking of different models on each dataset in all the experiments (The lowest rank is the best)}
    \label{fig:Rank}
\end{figure}

\subsection{Interpretable Meta-Model for best model predictions}
Based on our empirical study, we trained a decision tree of a maximum depth of 6 levels and a minimum impurity split of 0.3 to keep it simple and interpretable. This tree model is used to predict the best outperforming model based on dataset characteristics. We used all the experiments that we have carried out in our study to construct a new tabular dataset. The features listed in this dataset include the number of sessions, average session length, and average item frequency in both training and testing sets. We set the target variable as the best performing model out of the whole list of the evaluated models according to the $MRR@5$ evaluation metric. These models are distributed as 14, 36, 7, 4, 10, 5, and 10 instances for the S-POP, VSKNN, NARM, STAMP, NextItNet, SRGNN, and CSRM, respectively. On the other hand, the remaining models did not outperform in all the data splits used in our study. 
Our dataset was divided into ten cross-fold training and hold-out splits. The same decision tree was fitted to each training split to achieve an average accuracy of 87.17\% and 87.5\% on the training and hold-out splits, respectively. The visualization of one of these fitted trees, and the class distribution in the dataset can be found in Figure \ref{fig:tree} in the appendix. 

The most important features used in determining the outperforming model turn out to be in the following order: the average item frequency in the training set, the average session length in the testing set, the number of sessions in the training set, and the number of items in the training set. This simple tree model supports our previous findings of how different dataset characteristics can affect the performance of different models, and choosing the best one. In practice, such interpretable models can help the user to shorten the list of the models that are more likely to perform well given the characteristics of the given dataset. Besides, they can help assign weights for different models' predictions if an ensemble of multiple models is used for recommendation using the recommendations corresponding to each model. This experiment shows the potential in finding similar interpretable models that help in developing rules that guide the user to choose the suitable models for a specific dataset.

Our study suggests that using different models according to the different datasets' characteristics could lead to a better performance in the session-based recommendation task. Similar approaches to our decision-tree meta-model can predict which models will perform well with different dataset properties. This information can help combine the predictions out of multiple candidate models, which will consequently improve the final set of recommended items. Our dataset can also be extended easily with more e-commerce datasets that can increase the meta-model accuracy and reliability of predicting the best models.

\subsection{Overall Performance}
To judge the overall performance of the different models, we used a box-plot in Figure \ref{fig:Rank} to summarize the ranks of the examined models along with the evaluated datasets. Each chart represents a comparison among the ranks of the models in all the experiments related to one particular dataset. The model with the best performance (highest $HR@5$/$MRR@5$) takes a ranking of one, and the one with the worst performance takes a ranking of twelve. In general, {NARM}, {SRGNN}, and {CSRM} are the top three neural-based models in terms of both $HR@5$ and $MRR@5$ in all datasets. {VSKNN} has a good performance in {ROCKET} dataset with the smallest average session length among all datasets. In contrast with previous studies \citep*{ludewig2019empirical, ludewig2018evaluation}, VSKNN has worse performance than expected. When we investigated the reasons for this performance impairment, we found out that the preprocessing steps carried out in our study by removing consecutive clicks on the same items, keeping the items of low frequency, and the different evaluation procedures are the main reasons for the differences from these studies. {S-POP} has the best performance in {TMALL} dataset with the largest number of items and average session length. {NextItNet} has a good performance only in {RECSYS} dataset with the smallest number of items and largest average item frequency in the training set, which means that most items are well-represented in training set by many times. In Table \ref{tab:all} in the appendix, the different metrics are evaluated for $1,5,20$ predictions cut-off thresholds. Overall, the performance of neural-based models has been greatly improved with the new different architectures that emerged in the session-based recommendation. This improvement can also be observed when comparing the performance of neural-based models in older studies \citep*{jannach2017recurrent, ludewig2018evaluation} compared with more recent ones \citep*{ludewig2019empirical}. Hence, neural-based models have a comparable accuracy performance to the most developed nearest neighbor algorithms, and yet more research is needed to further extend these models.

When comparing the results obtained from this study with previous benchmarking studies like \citep*{jannach2017recurrent, ludewig2018evaluation, ludewig2019empirical}, it is observed that the relative overall performance of the evaluated methods on the whole datasets is not always the same. Also, the performance of the models among these studies changes by doing slightly different preprocessing steps. For example, the performance of the STAMP model drops considerably in \citep*{wu2019session} compared with the reported performance in \citep*{liu2018stamp} on the same evaluated datasets (RECSYS and CIKMCUP). For this reason, we find it difficult to draw general conclusions about the relative performance of the evaluated models. Additionally, there are very few publicly available real-world datasets in e-commerce. Hence, this suggests that understanding every single model's performance concerning the different dataset characteristics could provide the user with more insights to help him select the model that suits his dataset better in an interpretable way. Having a large number of real-world datasets covering the whole space of these characteristics sounds practically impossible. Thus, here in this study, we rely on creating artificially altered data splits out of the original datasets that could better understand the performance of the evaluated models.

\section{Conclusion}
\label{sec:6}
\subsection{Main Insights}
In this study, we investigated the current state-of-the-art neural-based models in addition to other baseline algorithms for the session-based recommendation task. Different experiments have been carried out trying to answer a set of research questions covering different characteristics of the evaluated datasets, in the e-commerce domain, during both training and testing phases. We used different evaluation metrics covering the accuracy of the models' recommendations, the coverage of predicted items, and their average popularity. Additionally, the consumption of computational resources during training and inference has been discussed in terms of the suitability to real-life e-commerce portals.

In general, neural-based models with attention mechanisms like {NARM} and {CSRM} in addition to recurrent models like {GRU4Rec+}, and the simple VSKNN algorithm, are the top-performing models over the majority of datasets with different characteristics. Besides, the neural-based models are characterized by having reasonable training time budgets and real-time processing during inference. Our results suggest that the training data recency and sizes have an observable effect on the prediction accuracy during inference. In e-commerce, it is clear that dynamic time modeling is a crucial part that needs to be further investigated and included in session-based algorithms to model general trends through different periods. Additionally, dataset characteristics like average session length, average item frequency, and the total number of sessions do have an impact on the models' performance.

Baseline models like nearest neighbors are still outperforming all other models when having relatively small training sizes or short sessions. Additionally, most of the models' performance degrades slowly on very long sessions, which suggests the need to improve the models' performance in these cases and accurately detect drifts in the user's preferences while making use of older events in the session efficiently.
In some cases, baseline algorithms outperform neural models; however, due to the computational complexity of these algorithms, especially during the inference time, neural-based models are preferable in making real-time recommendations.

\subsection{Challenges and Future Work}
Despite the recent leap achieved in improving the performance of neural-based methods in the session-based recommendation, many challenges still need to be tackled with new solutions. As future work, we suggest the following research points that help the community in understanding the current models better, and tackling these challenges with new solutions:
\begin{itemize}
    \item E-commerce domain is usually characterized by frequent changes in items properties. For example, sale campaigns on some items can affect the users' interest heavily. In addition, temporal changes like weather changes in different seasons, and trends in fashion items can also lead to significant drifts in users' preferences. Thus, it is quite important to start looking for models that can deal with possibly different types of items attributes, either they are nominal, numerical, or categorical, to improve the prediction accuracy. Additionally, temporal changes in these attributes should be taken into account while predicting different items. The possible effects of these trends were previously analyzed using an e-commerce use case \citep*{jannach2017session}. However, this is little explored in the literature due to the lack of publicly available datasets, including enough relevant information. So, more effort should be made to collect and publish such types of datasets that can help the research community better analyze these trends with different domains and a wide range of scenarios.
    \item Most of the current solutions require unique item identifiers to be used during training and prediction phases. However, in many domains having a fixed set of items is not a feasible solution. For example, new items can be added, and others can run out of stock in the e-commerce domain. So, training new models can be a tedious solution, especially for large datasets. Besides, models can suffer severely from the cold-start problem for those recently added items. Research work needs to investigate how to use the concept of the dynamic item embedding \citep*{kumar2019predicting} to be utilized in both the training and inferences phases instead of using a fixed set of unique identifiers.
    \item The current session-based recommendation systems do not take into account the different user interactions made during the session. On the one hand, different events like item view, add-to-cart, and add to wish list show different levels of interest from the user towards the items. On the other hand, other interactions can account for drifts in the user preferences like remove-from-cart and remove-from-wishlist. We believe that modeling such kinds of different interactions in a general way can lead to an improvement in the session-based recommendation.
    \item Although tuning the models hyper-parameters can be computationally expensive, it is quite important for further studies to perform extensive experiments on the most promising models to investigate the effect of changing different architecture hyper-parameters. Previous studies show that it is always the case that few hyper-parameters have a significant impact on the models' performance \citep*{hutter2007boosting,hutter2019automated}. As a candidate solution to reduce the search space to be investigated, many studies have introduced solutions to automated machine learning, including the neural network architecture search and hyperparameter optimization, which help carry out these studies more fairly and efficiently \citep*{elshawi2019automated}.
    \item Extensive evaluation of deep learning approaches in session-based recommendation in domains other than e-commerce like music playlist recommendations is not investigated yet. As different domains characteristics can affect the properties of data collected, and the performance of different models, it is quite important to answer similar research questions to the ones investigated here in other domains too. Besides, other evaluation metrics could also be computed to reveal new interesting information about the behavior of the models and the diversity of their predictions \citep*{shani2011evaluating}. For example, the Gini index could be evaluated for the models' recommendations to understand if the predictions are biased towards some items than the others.
    \item In session-based recommendation, it is typical to train a model using the sessions collected during a specific period and evaluating the model using the sessions collected in the subsequent days to that period. Although this approach was always followed in our study, it is also possible to experiment using different train-test splits in some experiments like RQ4, RQ5, and RQ6, where a random set of sessions where chosen from the entire list of the existing sessions. One of the limitations of this work is that we used only a single train-test split due to our experiments' computational complexity. Hence, as a future work, evaluating the results of some research questions using multiple train‐test splits can be used to confirm and generalize our main conclusions.
\end{itemize}

\begin{acknowledgements}
This work is funded by Rakuten, Inc. (Grant VLTTI19503). The work has also been partially supported by the Estonian Centre of Excellence in IT (EXCITE) funded by the European Regional Development Fund and the Rakuten, Inc. (Grant VLTTI19503). The authors also gratefully acknowledge the support of NVIDIA Corporation with the donation of the Titan X Pascal GPU.
\end{acknowledgements}

\bibliographystyle{spbasic}
\bibliography{biblio}

\clearpage
\section{Appendix}
\subsection{Properties of the datasets in all the experiments}
\begin{table*}[!h]
\centering
\caption{Final Statistics of datasets splits used in all the different evaluation Experiments}
\resizebox{\textwidth}{!}{
\label{tab:appendixtab0}
\arrayrulecolor{black}
\begin{tabular}{|c|c|c|c|c|c|c|c|c|c|c|c|c|c|} 
\hline
{\cellcolor[rgb]{0.753,0.753,0.753}} & \multirow{3}{*}{} & \multicolumn{6}{c|}{{\cellcolor[rgb]{0.871,0.871,0.871}} \textbf{RECSYS} } & \multicolumn{6}{c|}{{\cellcolor[rgb]{0.871,0.871,0.871}} \textbf{CIKMCUP} } \\ 
\hhline{|>{\arrayrulecolor[rgb]{0.753,0.753,0.753}}-~>{\arrayrulecolor{black}}------------|}
{\cellcolor[rgb]{0.753,0.753,0.753}} &  & \multicolumn{3}{c|}{ \textbf{Training Set} } & \multicolumn{3}{c|}{ \textbf{Test Set} } & \multicolumn{3}{c|}{ \textbf{Training Set} } & \multicolumn{3}{c|}{ \textbf{Test Set} } \\ 
\hhline{|>{\arrayrulecolor[rgb]{0.753,0.753,0.753}}-~>{\arrayrulecolor{black}}------------|}
\multirow{-3}{*}{{\cellcolor[rgb]{0.753,0.753,0.753}}\begin{tabular}[c]{@{}>{\cellcolor[rgb]{0.753,0.753,0.753}}c@{}}\textbf{Target }\\\textbf{RQ} \end{tabular}} &  & \begin{tabular}[c]{@{}c@{}}\textbf{No. of}\\\textbf{Sessions}\end{tabular} & \begin{tabular}[c]{@{}c@{}}\textbf{Avg. }\\\textbf{ Session }\\\textbf{ Length} \end{tabular} & \begin{tabular}[c]{@{}c@{}}\textbf{Avg. }\\\textbf{ Item }\\\textbf{ Freq.} \end{tabular} & \begin{tabular}[c]{@{}c@{}}\textbf{No. of}\\\textbf{Sessions}\end{tabular} & \begin{tabular}[c]{@{}c@{}}\textbf{Avg. }\\\textbf{ Session}\\\textbf{ Length} \end{tabular} & \begin{tabular}[c]{@{}c@{}}\textbf{Avg.}\\\textbf{Item}\\\textbf{Freq.} \end{tabular} & \begin{tabular}[c]{@{}c@{}}\textbf{No. of}\\\textbf{Sessions}\end{tabular} & \begin{tabular}[c]{@{}c@{}}\textbf{Avg. }\\\textbf{ Session }\\\textbf{ Length} \end{tabular} & \begin{tabular}[c]{@{}c@{}}\textbf{Avg. }\\\textbf{ Item }\\\textbf{ Freq. }\end{tabular} & \begin{tabular}[c]{@{}c@{}}\textbf{No. of}\\\textbf{Sessions}\end{tabular} & \begin{tabular}[c]{@{}c@{}}\textbf{Avg. }\\\textbf{ Session}\\\textbf{ Length} \end{tabular} & \begin{tabular}[c]{@{}c@{}}\textbf{Avg.}\\\textbf{Item}\\\textbf{Freq.} \end{tabular} \\ 
\hline
\multirow{3}{*}{\textbf{RQ1} } & \textbf{Long Sessions}  & 447K & 13.82 & 187.06 & 13K & 4.58 & 1000 &    26K & 13.14 & 4.33 &    2K & 4.80 & 20.95 \\ 
\cline{2-14}
 & \textbf{Intermediate Sessions}  & 1402K & 5.59 & 229.87 & 13K & 4.58 & 1336 & 66K~ & 6.05 & 5.04 & 2K & 4.76 & 28.28 \\ 
\cline{2-14}
 & \textbf{Short Sessions}  & 4888K & 2.53 & 336.73 & 13K & 4.58 & 2184 & 116K & 2.71 & 4.38 & 2K & 4.66 & 33.07 \\ 
\hline
\multirow{3}{*}{\textbf{RQ2} } & \textbf{Short Sessions}  & 145K & 4.15 & 31.67 &    9K & 2.57 & 1059 & 62K & 5.04 & 4.67 &    1.2K & 2.64 & 31.21 \\ 
\cline{2-14}
 & \textbf{Intermediate Sessions}  & 145K & 4.15 & 31.67 & 3K & 5.77 & 889.1 & 62K & 5.04 & 4.67 & 716 & 5.53 & 32.50 \\ 
\cline{2-14}
 & \textbf{Long Sessions}  & 145K & 4.15 & 31.67 & 1.4K & 14.38 & 669.7 & 62K & 5.04 & 4.67 & 299 & 11.25 & 30.00 \\ 
\hline
\multirow{4}{*}{\textbf{RQ3} } & \textbf{Very Low Freq.}  & 145K & 4.15 & 31.67 &    13K & 4.05 & 19.89 & 62K & 5.04 & 4.67 & 2K & 3.19 & 4.09 \\ 
\cline{2-14}
 & \textbf{Low Freq.}  & 145K & 4.15 & 31.67 & 13K & 3.80 & 36.96 & 62K & 5.04 & 4.67 & 2K & 2.10 & 9.54 \\ 
\cline{2-14}
 & \textbf{Intermediate Freq.}  & 145K & 4.15 & 31.67 & 13K & 3.44 & 71.10 & 62K & 5.04 & 4.67 & 2K & 1.56 & 15.15 \\ 
\cline{2-14}
 & \textbf{High Freq.}  & 145K & 4.15 & 31.67 & 13K & 3.17 & \begin{tabular}[c]{@{}c@{}}104.86\\ \end{tabular} & 62K & 5.04 & 4.67 & 2K & 1.27 & 20.31 \\ 
\hline
\multirow{3}{*}{\textbf{RQ4} } & \textbf{Mixed}  & 743K & 3.97 & 94.88 & 13K & 4.58 & 1453 & 18K & 4.97 & 2.49 & 1.9K & 4.47 & 14.80 \\ 
\cline{2-14}
 & \textbf{Recent}  & 470K & 4.02 & 79.19 & 13K & 4.58 & 1602 &    23K & 5.03 & 2.99 & 2K & 4.52 & 11.18 \\ 
\cline{2-14}
 & \textbf{Old}  & 662K & 3.91 & 100.17 & 4.5K & 3.45 & 519.7 & 22K & 5.00 & 3.01 & 1.4K & 3.81 & 10.39 \\ 
\hline
\multirow{4}{*}{\textbf{RQ5} } & \textbf{Large Portion}  &    842K & 3.91 & 99.60 &    13K & 4.57 & 564.3 &    104K & 5.06 & 5.64 &    2K & 4.85 & 39.23 \\ 
\cline{2-14}
 & \textbf{Medium Portion}  & 421K & 3.91 & 55.49 & 13K & 4.57 & 282.6 & 26K & 5.08 & 2.75 & 2K & 4.47 & 11.89 \\ 
\cline{2-14}
 & \textbf{Small Portion}  & 105K & 3.93 & 18.81 & 13K & 4.54 & 72.63 & 13K & 5.08 & 2.06 & 1.8K & 4.17 & 7.05 \\ 
\cline{2-14}
 & \textbf{Very Small Portion}  & 26K & 3.89 & 7.35 & 12K & 4.16 & 21.78 & 3K & 5.13 & 1.37 & 1.2K & 3.53 & 3.12 \\ 
\hline
\multirow{4}{*}{\textbf{RQ6} } & \textbf{2 days}  & 145K & 4.15 & 31.67 &    13K & 4.57 & 852.8 & 5K & 5.01 & 1.74 &    1.6K & 4.04 & 5.18 \\ 
\cline{2-14}
 & \textbf{7 days}  &    354K & 4.1 & 102.91 & 13K & 4.58 & 1437 & 16K & 5.01 & 2.59 & 2K & 4.44 & 11.40 \\ 
\cline{2-14}
 & \textbf{14 days}  & 635K & 4.04 & 63.21 & 13K & 4.58 & 1835 & 28K & 4.98 & 3.25 & 2K & 4.59 & 17.21 \\ 
\cline{2-14}
 & \textbf{30 days}  & 1241K & 4.05 & 174.07 & 13K & 4.58 & 2472 & 62K & 5.04 & 4.67 & 2K & 4.77 & 31.27 \\ 
\hline
\textbf{RQ7}  &  & 421K & 3.91 & 55.49 & 13K & 4.57 & 282.6 & 207K & 5.07 & 8.57 & 2K & 4.93 & 75.20 \\ 
\hline
\textbf{RQ8}  &  & 421K & 3.91 & 55.49 & 13K & 4.57 & 282.6 & 207K & 5.07 & 8.57 & 2K & 4.93 & 75.20 \\ 
\hline
{\cellcolor[rgb]{0.753,0.753,0.753}} & \multirow{3}{*}{} & \multicolumn{6}{c|}{{\cellcolor[rgb]{0.871,0.871,0.871}} \textbf{TMALL} } & \multicolumn{6}{c|}{{\cellcolor[rgb]{0.871,0.871,0.871}} \textbf{ROCKET} } \\ 
\hhline{|>{\arrayrulecolor[rgb]{0.753,0.753,0.753}}-~>{\arrayrulecolor{black}}------------|}
{\cellcolor[rgb]{0.753,0.753,0.753}} &  & \multicolumn{3}{c|}{ \textbf{Training Set} } & \multicolumn{3}{c|}{ \textbf{Test Set} } & \multicolumn{3}{c|}{ \textbf{Training Set} } & \multicolumn{3}{c|}{ \textbf{Test Set} } \\ 
\hhline{|>{\arrayrulecolor[rgb]{0.753,0.753,0.753}}-~>{\arrayrulecolor{black}}------------|}
\multirow{-3}{*}{{\cellcolor[rgb]{0.753,0.753,0.753}}} &  & \begin{tabular}[c]{@{}c@{}}\textbf{No. of}\\\textbf{Sessions}\end{tabular} & \begin{tabular}[c]{@{}c@{}}\textbf{Avg. }\\\textbf{ Session }\\\textbf{ Length} \end{tabular} & \begin{tabular}[c]{@{}c@{}}\textbf{Avg. }\\\textbf{ Item }\\\textbf{ Freq.} \end{tabular} & \begin{tabular}[c]{@{}c@{}}\textbf{No. of}\\\textbf{Sessions}\end{tabular} & \begin{tabular}[c]{@{}c@{}}\textbf{Avg. }\\\textbf{ Session}\\\textbf{ Length} \end{tabular} & \begin{tabular}[c]{@{}c@{}}\textbf{Avg.}\\\textbf{Item}\\\textbf{Freq.} \end{tabular} & \begin{tabular}[c]{@{}c@{}}\textbf{No. of}\\\textbf{Sessions}\end{tabular} & \begin{tabular}[c]{@{}c@{}}\textbf{Avg. }\\\textbf{ Session }\\\textbf{ Length} \end{tabular} & \begin{tabular}[c]{@{}c@{}}\textbf{Avg. }\\\textbf{ Item }\\\textbf{ Freq.} \end{tabular} & \begin{tabular}[c]{@{}c@{}}\textbf{No. of}\\\textbf{Sessions}\end{tabular} & \begin{tabular}[c]{@{}c@{}}\textbf{Avg. }\\\textbf{ Session}\\\textbf{ Length} \end{tabular} & \begin{tabular}[c]{@{}c@{}}\textbf{Avg.}\\\textbf{Item}\\\textbf{Freq.} \end{tabular} \\ 
\hline
\multirow{3}{*}{\textbf{RQ1} } & \textbf{Long Sessions}  & 488K & 12.79 & 11.82 & 9K & 9.72 & 143.9 & 13K & 16.73 & 4.2 & 1.3K & 3.93 & 17.49 \\ 
\cline{2-14}
 & \textbf{Intermediate Sessions}  &    462K & 5.35 & 7.26 & 9K & 9.03 & 80.10 & 41K & 5.53 & 4.32 & 1.4K & 3.72 & 17.76 \\ 
\cline{2-14}
 & \textbf{Short Sessions}  & 627K & 2.6 & 6.05 & 9K & 8.56 & 70.97 & 205K & 2.43 & 4.33 & 1.8K & 3.60 & 36.42 \\ 
\hline
\multirow{3}{*}{\textbf{RQ2} } & \textbf{Short Sessions}  & 71K & 7.86 & 3.81 & 3K & 2.59 & 42.05 & 51K & 3.47 & 3.17 & 1.2K & 2.38 & 26.07 \\ 
\cline{2-14}
 & \textbf{Intermediate Sessions}  & 71K & 7.86 & 3.81 & 3K & 5.21 & 36.92 & 51K & 3.47 & 3.17 & 260 & 5.14 & 20.00 \\ 
\cline{2-14}
 & \textbf{Long Sessions}  & 71K & 7.86 & 3.81 & 3K & 17.43 & 25.92 & 51K & 3.47 & 3.17 & 96 & 16.5 & 18.72 \\ 
\hline
\multirow{4}{*}{\textbf{RQ3} } & \textbf{Very Low Freq.}  & 71K & 7.86 & 3.81 & 9K & 2.12 & 11.06 & 259K & 3.65 & 7.03 & 1.8K & 3.03 & 4.78 \\ 
\cline{2-14}
 & \textbf{Low Freq.}  & 71K & 7.86 & 3.81 & 9K & 1.52 & 15.99 & 259K & 3.65 & 7.03 & 1.8K & 2.23 & 12.12 \\ 
\cline{2-14}
 & \textbf{Intermediate Freq.}  & 71K & 7.86 & 3.81 & 9K & 1.18 & 21.67 & 259K & 3.65 & 7.03 & 1.8K & 1.73 & 19.85 \\ 
\cline{2-14}
 & \textbf{High Freq.}  & 71K & 7.86 & 3.81 & 9K & 1.08 & 24.76 & 259K & 3.65 & 7.03 & 1.8K & 1.44 & 27.36 \\ 
\hline
\multirow{3}{*}{\textbf{RQ4} } & \textbf{Mixed}  & 265K & 7.44 & 6.1 & 9K & 9.39 & 63.09 & 37K & 3.6 & 2.71 & 1.4K & 3.66 & 14.31 \\ 
\cline{2-14}
 & \textbf{Recent}  & 327K & 7.25 & 7.62 & 9K & 9.52 & 91.46 & 34K & 3.47 & 2.72 & 1.4K & 3.7 & 17.07 \\ 
\cline{2-14}
 & \textbf{Old}  & 231K & 7.14 & 6.52 & 7K & 6.53 & 50.95 & 42K & 3.72 & 3.18 & 1K & 3.36 & 11.28 \\ 
\hline
\multirow{4}{*}{\textbf{RQ5} } & \begin{tabular}[c]{@{}c@{}}\textbf{Large}\\\textbf{ Portion} \end{tabular} & 186K & 7.03 & 4.92 & 9K & 8.76 & 40.27 & 130K & 3.66 & 4.77 & 1.7K & 3.68 & 33.80 \\ 
\cline{2-14}
 & \textbf{Medium Portion}  & 91K & 7.16 & 3.58 & 8.5K & 8.06 & 22.93 & 32K & 3.68 & 2.46 & 1.3K & 3.59 & 10.33 \\ 
\cline{2-14}
 & \textbf{Small Portion}  & 23K & 7.26 & 2.17 & 7K & 6.5 & 8.22 & 16K & 3.67 & 1.89 & 1K & 3.5 & 6.26 \\ 
\cline{2-14}
 & \textbf{Very Small Portion}  & 5.5K & 7.22 & 1.56 & 5K & 4.82 & 3.54 & 4K & 3.61 & 1.31 & 512 & 3.11 & 2.95 \\ 
\hline
\multirow{4}{*}{\textbf{RQ6} } & \textbf{2 days}  & 71K & 7.86 & 3.81 & 9K & 8.61 & 28.94 & 3.8K & 3.41 & 1.46 & 674 & 3.23 & 5.51 \\ 
\cline{2-14}
 & \textbf{7 days}  & 228K & 7.46 & 6.42 & 9K & 9.37 & 69.88 & 13K & 3.49 & 1.97 & 1.1K & 3.61 & 9.70 \\ 
\cline{2-14}
 & \textbf{14 days}  & 462K & 6.95 & 8.98 & 9K & 9.62 & 123.2 & 24K & 3.47 & 2.41 & 1.3K & 3.68 & 13.68 \\ 
\cline{2-14}
 & \textbf{30 days}  & 883K & 6.42 & 12.02 & 9K & 9.76 & 184.4 & 51K & 3.47 & 3.17 & 1.6K & 3.72 & 21.92 \\ 
\hline
\textbf{RQ7}  &  & 91K & 7.16 & 3.58 & 8.5K & 8.06 & 22.93 & 259K & 3.65 & 7.03 & 1.8K & 3.68 & 64.43 \\ 
\hline
\textbf{RQ8}  &  & 91K & 7.16 & 3.58 & 8.5K & 8.06 & 22.93 & 259K & 3.65 & 7.03 & 1.8K & 3.68 & 64.43 \\
\hline
\end{tabular}}
\end{table*}

\clearpage
\subsection{Experiments Results}

\begin{table*}[!h]
\centering
\caption{RQ1: Effect of using different training session length on algorithms performance.}
\label{tab:rq1}
\arrayrulecolor{black}
\resizebox{\textwidth}{!}
{
\begin{tabular}{|c|ccc|ccc|ccc|ccc|} 
\hhline{>{\arrayrulecolor[rgb]{0.149,0.149,0.149}}->{\arrayrulecolor{black}}------------|}
\multicolumn{1}{l|}{{\cellcolor[rgb]{0.149,0.149,0.149}}} & \multicolumn{6}{c|}{\textbf{RECSYS}} & \multicolumn{6}{c|}{\textbf{CIKMCUP}} \\ 
\hline
{\cellcolor[rgb]{0.149,0.149,0.149}} & \multicolumn{3}{c|}{\textbf{HR@5}} & \multicolumn{3}{c|}{\textbf{MRR@5}} & \multicolumn{3}{c|}{\textbf{HR@5}} & \multicolumn{3}{c|}{\textbf{MRR@5}} \\ 
\hline
\multicolumn{1}{l}{{\cellcolor[rgb]{0.149,0.149,0.149}}} & \textbf{Short} & \textbf{Intermediate} & \textbf{Long} & \textbf{Short} & \textbf{Intermediate} & \textbf{Long} & \textbf{Short} & \textbf{Intermediate} & \textbf{Long} & \textbf{Short} & \textbf{Intermediate} & \textbf{Long} \\ 
\cline{2-13}
\textbf{ SPOP} & 0.12266 & 0.12276 & 0.12179 & 0.06912 & 0.0686 & 0.068 & 0.12132 & 0.11757 & 0.11394 & 0.07221 & 0.07001 & 0.06793 \\ 
\textbf{ AR} & 0.2579 & 0.27188 & 0.26112 & 0.15759 & 0.16427 & 0.15784 & 0.12547 & 0.1388 & 0.127 & 0.07238 & 0.07587 & 0.06782 \\ 
\textbf{ SR} & 0.24833 & 0.27026 & 0.26726 & 0.15295 & 0.16194 & 0.16272 & 0.10735 & 0.12339 & 0.1169 & 0.06262 & 0.06919 & 0.06181 \\ 
\textbf{ VSKNN} & 0.14516 & 0.0834 & 0.24868 & 0.09956 & 0.05627 & 0.15752 & 0.12081 & 0.13116 & 0.14978 & 0.07924 & 0.07936 & 0.08961 \\ 
\textbf{ SMF} & 0.28226 & 0.29364 & 0.27855 & 0.14128 & 0.14973 & 0.13823 & 0.13326 & 0.14014 & 0.1307 & 0.06803 & 0.07067 & 0.06515 \\
\hline
\end{tabular}}
\end{table*}


\begin{table*}[!h]
\centering
\caption{RQ2: Performance of the models on different testing session lengths.}
\label{tab:rq2}
\arrayrulecolor{black}
\resizebox{\textwidth}{!}
{
\begin{tabular}{|c|ccc|ccc|ccc|ccc|} 
\hhline{>{\arrayrulecolor[rgb]{0.149,0.149,0.149}}->{\arrayrulecolor{black}}------------|}
\multicolumn{1}{l|}{{\cellcolor[rgb]{0.149,0.149,0.149}}} & \multicolumn{6}{c|}{\textbf{RECSYS}} & \multicolumn{6}{c|}{\textbf{CIKMCUP}} \\ 
\hline
{\cellcolor[rgb]{0.149,0.149,0.149}} & \multicolumn{3}{c|}{\textbf{HR@5}} & \multicolumn{3}{c|}{\textbf{MRR@5}} & \multicolumn{3}{c|}{\textbf{HR@5}} & \multicolumn{3}{c|}{\textbf{MRR@5}} \\ 
\hline
\multicolumn{1}{l}{{\cellcolor[rgb]{0.149,0.149,0.149}}} & \textbf{Short} & \textbf{Intermediate} & \textbf{Long} & \textbf{Short} & \textbf{Intermediate} & \textbf{Long} & \textbf{Short} & \textbf{Intermediate} & \textbf{Long} & \textbf{Short} & \textbf{Intermediate} & \textbf{Long} \\ 
\cline{2-13}
\textbf{ SPOP} & 0.07384 & 0.14258 & 0.1636 & 0.04449 & 0.08335 & 0.08591 & 0.06207 & 0.11427 & 0.15337 & 0.04385 & 0.07187 & 0.08378 \\ 
\textbf{ AR} & 0.38767 & 0.30338 & 0.21566 & 0.25205 & 0.18148 & 0.12749 & 0.18727 & 0.15503 & 0.11187 & 0.10216 & 0.08386 & 0.06085 \\ 
\textbf{ SR} & 0.39096 & 0.30644 & 0.21935 & 0.25272 & 0.18307 & 0.13043 & 0.17307 & 0.13897 & 0.10409 & 0.09832 & 0.07778 & 0.05749 \\ 
\textbf{ VSKNN} & 0.36927 & 0.29541 & 0.2387 & 0.23924 & 0.17427 & 0.13559 & 0.18488 & 0.1607 & 0.1648 & 0.11907 & 0.09674 & 0.09222 \\ 
\textbf{ SMF} & 0.34652 & 0.29268 & 0.24241 & 0.18395 & 0.15282 & 0.12526 & 0.16446 & 0.16972 & 0.13466 & 0.08197 & 0.08744 & 0.06645 \\
\hline
\end{tabular}}
\end{table*}

\newpage

\begin{table*}[!h]
\centering
\caption{RQ3: Performance of different models using different items frequency values in training set ($<50,<100,<200,<300$) for RECSYS and TMALL and ($<10,<30,<60,<100$) for CIKMCUP and ROCKET.}
\label{tab:rq3}
\arrayrulecolor{black}
\resizebox{\textwidth}{!}
{
\begin{tabular}{|c|clcc|clcc|clcc|clcc|} 
\hline
\multicolumn{1}{l|}{{\cellcolor[rgb]{0.149,0.149,0.149}}} & \multicolumn{8}{c|}{\textbf{RECSYS} } & \multicolumn{8}{c|}{\textbf{CIKMCUP} } \\ 
\hline
{\cellcolor[rgb]{0.149,0.149,0.149}} & \multicolumn{4}{c|}{\textbf{HR@5} } & \multicolumn{4}{c|}{\textbf{MRR@5} } & \multicolumn{4}{c|}{\textbf{HR@5} } & \multicolumn{4}{c|}{\textbf{MRR@5} } \\ 
\hline
\multicolumn{1}{l}{{\cellcolor[rgb]{0.149,0.149,0.149}}} & \textbf{$<50(<10)$}  & \multicolumn{1}{c}{\textbf{$<100(<30)$}} & \textbf{$<200(<60)$}  & \textbf{$<300(<100)$}  & \textbf{$<50$}  & \multicolumn{1}{c}{\textbf{$<100$}} & \textbf{$<200$}  & \textbf{$<300$}  & \textbf{$<50$}  & \multicolumn{1}{c}{\textbf{$<100$}} & \textbf{$<200$}  & \textbf{$<300$}  & \textbf{$<50$}  & \multicolumn{1}{c}{\textbf{$<100$}} & \textbf{$<200$}  & \textbf{$<300$}  \\ 
\cline{2-17}
\textbf{ SPOP} & 0.09101 & 0.09597 & 0.10111 & 0.10068 & 0.05233 & 0.0544 & 0.0577 & 0.05716 & 0.06932 & 0.09455 & 0.1079 & 0.1137 & 0.04055 & 0.0565 & 0.06404 & 0.06799 \\ 
\textbf{ AR} & 0.1443 & 0.16365 & 0.17659 & 0.18055 & 0.08025 & 0.0949 & 0.10249 & 0.10354 & 0.05657 & 0.09687 & 0.11444 & 0.12731 & 0.02686 & 0.04963 & 0.0604 & 0.06803 \\ 
\textbf{ SR} & 0.14384 & 0.16941 & 0.18121 & 0.18478 & 0.0809 & 0.09716 & 0.10548 & 0.10658 & 0.05284 & 0.08061 & 0.09772 & 0.10986 & 0.02869 & 0.04493 & 0.05465 & 0.06181 \\ 
\textbf{ VSKNN} & 0.24221 & 0.22056 & 0.22025 & 0.22658 & 0.17753 & 0.1581 & 0.15794 & 0.15997 & 0.17844 & 0.14333 & 0.13052 & 0.12706 & 0.12518 & 0.09318 & 0.08164 & 0.07717 \\ 
\textbf{ SMF} & 0.20131 & 0.24493 & 0.26802 & 0.26617 & 0.10311 & 0.12805 & 0.14179 & 0.14373 & 0.05227 & 0.11041 & 0.12816 & 0.13817 & 0.02685 & 0.05297 & 0.0617 & 0.07005 \\
\hline
\end{tabular}}
\end{table*}


\begin{table*}[!h]
\centering
\caption{RQ4: Effect of data recency on models performance for different datasets.}
\label{tab:rq4}
\arrayrulecolor{black}
\resizebox{\textwidth}{!}
{
\begin{tabular}{|c|ccc|ccc|ccc|ccc|} 
\hhline{>{\arrayrulecolor[rgb]{0.149,0.149,0.149}}->{\arrayrulecolor{black}}------------|}
\multicolumn{1}{l|}{{\cellcolor[rgb]{0.149,0.149,0.149}}} & \multicolumn{6}{c|}{\textbf{RECSYS}} & \multicolumn{6}{c|}{\textbf{CIKMCUP}} \\ 
\hline
{\cellcolor[rgb]{0.149,0.149,0.149}} & \multicolumn{3}{c|}{\textbf{HR@5}} & \multicolumn{3}{c|}{\textbf{MRR@5}} & \multicolumn{3}{c|}{\textbf{HR@5}} & \multicolumn{3}{c|}{\textbf{MRR@5}} \\ 
\hline
\multicolumn{1}{l}{{\cellcolor[rgb]{0.149,0.149,0.149}}} & \textbf{Recent} & \textbf{Old} & \textbf{Mixed} & \textbf{Recent} & \textbf{Old} & \textbf{Mixed} & \textbf{Recent} & \textbf{Old} & \textbf{Mixed} & \textbf{Recent} & \textbf{Old} & \textbf{Mixed} \\ 
\cline{2-13}
\textbf{ SPOP} & 0.12481 & 0.13773 & 0.12352 & 0.07032 & 0.08607 & 0.06964 & 0.12274 & 0.12135 & 0.12213 & 0.07284 & 0.07215 & 0.07153 \\ 
\textbf{ AR} & 0.29245 & 0.1776 & 0.28489 & 0.1806 & 0.10443 & 0.17524 & 0.13427 & 0.11955 & 0.12539 & 0.07544 & 0.06687 & 0.0685 \\ 
\textbf{ SR} & 0.29712 & 0.18192 & 0.28927 & 0.18315 & 0.10445 & 0.1783 & 0.10969 & 0.09605 & 0.09827 & 0.06309 & 0.05371 & 0.05832 \\ 
\textbf{ VSKNN} & 0.25384 & 0.18167 & 0.24997 & 0.15025 & 0.13974 & 0.15719 & 0.15665 & 0.19082 & 0.17319 & 0.09448 & 0.12078 & 0.1034 \\ 
\textbf{ SMF} & 0.28946 & 0.21291 & 0.29022 & 0.14583 & 0.0957 & 0.14433 & 0.14018 & 0.12912 & 0.13678 & 0.0712 & 0.06884 & 0.07295 \\
\hline
\end{tabular}}
\end{table*}

\newpage

\begin{table*}[!h]
\centering
\caption{RQ5: The models' performance on different training set sizes.}
\label{tab:rq5}
\arrayrulecolor{black}
\resizebox{\textwidth}{!}
{
\begin{tabular}{|c|clcc|clcc|clcc|clcc|} 
\hline
\multicolumn{1}{l|}{{\cellcolor[rgb]{0.149,0.149,0.149}}} & \multicolumn{8}{c|}{\textbf{RECSYS} } & \multicolumn{8}{c|}{\textbf{CIKMCUP} } \\ 
\hline
{\cellcolor[rgb]{0.149,0.149,0.149}} & \multicolumn{4}{c|}{\textbf{HR@5} } & \multicolumn{4}{c|}{\textbf{MRR@5} } & \multicolumn{4}{c|}{\textbf{HR@5} } & \multicolumn{4}{c|}{\textbf{MRR@5} } \\ 
\hline
\multicolumn{1}{l}{{\cellcolor[rgb]{0.149,0.149,0.149}}} & \textbf{Large}  & \multicolumn{1}{c}{\textbf{Medium}} & \textbf{Small}  & \textbf{V.Small}  & \textbf{Large}  & \multicolumn{1}{c}{\textbf{Medium}} & \textbf{Small}  & \textbf{V.Small}  & \textbf{Large}  & \multicolumn{1}{c}{\textbf{Medium}} & \textbf{Small}  & \textbf{V.Small}  & \textbf{Large}  & \multicolumn{1}{c}{\textbf{Medium}} & \textbf{Small}  & \textbf{V.Small}  \\ 
\cline{2-17}
\textbf{ SPOP} & 0.123 & 0.23887 & 0.12375 & 0.13246 & 0.06885 & 0.17229 & 0.06909 & 0.07456 & 0.11678 & 0.12099 & 0.12219 & 0.13977 & 0.06926 & 0.07126 & 0.07293 & 0.08585 \\ 
\textbf{ AR} & 0.2539 & 0.20174 & 0.15408 & 0.17881 & 0.15159 & 0.12179 & 0.09584 & 0.11112 & 0.13897 & 0.11553 & 0.10946 & 0.09884 & 0.07758 & 0.0616 & 0.05932 & 0.05875 \\ 
\textbf{ SR} & 0.25294 & 0.23781 & 0.19775 & 0.14106 & 0.15411 & 0.14695 & 0.11946 & 0.08816 & 0.127 & 0.0957 & 0.08297 & 0.07022 & 0.07044 & 0.0536 & 0.04847 & 0.03977 \\ 
\textbf{ VSKNN} & 0.11868 & 0.18405 & 0.15433 & 0.19799 & 0.08256 & 0.11959 & 0.10386 & 0.12862 & 0.12383 & 0.14896 & 0.17333 & 0.26981 & 0.07731 & 0.09177 & 0.10744 & 0.16946 \\ 
\textbf{ SMF} & 0.26748 & 0.25059 & 0.1912 & 0.17083 & 0.13632 & 0.12911 & 0.09989 & 0.09317 & 0.14284 & 0.11634 & 0.11797 & 0.12938 & 0.07414 & 0.0622 & 0.05994 & 0.06912 \\
\hline
\end{tabular}}
\end{table*}


\begin{table}[!h]
\centering
\caption{RQ6: The models' performance on different training set time spans.}
\label{tab:rq6}
\arrayrulecolor{black}
\resizebox{\textwidth}{!}
{
\begin{tabular}{|c|clcc|clcc|clcc|clcc|} 
\hline
\multicolumn{1}{l|}{{\cellcolor[rgb]{0.149,0.149,0.149}}} & \multicolumn{8}{c|}{\textbf{RECSYS} } & \multicolumn{8}{c|}{\textbf{CIKMCUP} } \\ 
\hline
{\cellcolor[rgb]{0.149,0.149,0.149}} & \multicolumn{4}{c|}{\textbf{HR@5} } & \multicolumn{4}{c|}{\textbf{MRR@5} } & \multicolumn{4}{c|}{\textbf{HR@5} } & \multicolumn{4}{c|}{\textbf{MRR@5} } \\ 
\hline
\multicolumn{1}{l}{{\cellcolor[rgb]{0.149,0.149,0.149}}} & \textbf{2days}  & \multicolumn{1}{c}{\textbf{7d}} & \textbf{14d}  & \textbf{30d}  & \textbf{2d}  & \multicolumn{1}{c}{\textbf{7d}} & \textbf{14d}  & \textbf{30d}  & \textbf{2d}  & \multicolumn{1}{c}{\textbf{7d}} & \textbf{14d}  & \textbf{30d}  & \textbf{2d}  & \multicolumn{1}{c}{\textbf{7d}} & \textbf{14d}  & \textbf{30d}  \\ 
\cline{2-17}
\textbf{ SPOP} & 0.13353 & 0.12558 & 0.12507 & 0.12356 & 0.07388 & 0.07079 & 0.07026 & 0.0694 & 0.24145 & 0.2112 & 0.12168 & 0.11996 & 0.17616 & 0.14873 & 0.07216 & 0.071 \\ 
\textbf{ AR} & 0.29183 & 0.29309 & 0.2579 & 0.28838 & 0.18077 & 0.18059 & 0.15759 & 0.17865 & 0.13353 & 0.13191 & 0.13511 & 0.144 & 0.07294 & 0.07332 & 0.07576 & 0.07949 \\ 
\textbf{ SR} & 0.10015 & 0.11067 & 0.11818 & 0.12579 & 0.05643 & 0.06239 & 0.06317 & 0.07129 & 0.11615 & 0.12723 & 0.13432 & 0.14264 & 0.06567 & 0.07218 & 0.07174 & 0.08067 \\ 
\textbf{ VSKNN} & 0.30713 & 0.26796 & 0.23744 & 0.20827 & 0.17618 & 0.16437 & 0.16055 & 0.13191 & 0.23906 & 0.16883 & 0.14673 & 0.12945 & 0.14352 & 0.10287 & 0.08866 & 0.07838 \\ 
\textbf{ SMF} & 0.28808 & 0.2933 & 0.2893 & 0.29315 & 0.15163 & 0.15178 & 0.14742 & 0.14412 & 0.14554 & 0.14331 & 0.14086 & 0.14701 & 0.07703 & 0.07536 & 0.07296 & 0.07247 \\
\hline
\end{tabular}}
\end{table}

\newpage

\begin{table}[!t]
\centering
\caption{Training using random (1/16, All, 1/16, All) portions of the full training set for (RECSYS, CIKMCUP, TMALL, ROCKET) datasets respectively. Evaluation is done on the original testing sets. The HR, MRR, Coverage and Popularity metrics with different cut-off thresholds are reported. Highest performance is made in bold for each dataset.}
\label{tab:all}
\arrayrulecolor{black}
\resizebox{\textwidth}{!}{\begin{tabular}{|c|ccc|ccc|ccc|ccc|} 
\hline
{\cellcolor[rgb]{0.2,0.2,0.2}} & \multicolumn{3}{c|}{\textbf{HR@} } & \multicolumn{3}{c|}{\textbf{MRR@} } & \multicolumn{3}{c|}{\textbf{COV@} } & \multicolumn{3}{c|}{\textbf{POP@} } \\ 
\hhline{|>{\arrayrulecolor[rgb]{0.2,0.2,0.2}}->{\arrayrulecolor{black}}------------|}
\multirow{-2}{*}{{\cellcolor[rgb]{0.2,0.2,0.2}}\textcolor{white}{RECSYS}} & \textbf{1}  & \textbf{5}  & \textbf{20}  & \textbf{1}  & \textbf{5}  & \textbf{20}  & \textbf{1}  & \textbf{5}  & \textbf{20}  & \textbf{1}  & \textbf{5}  & \textbf{20}  \\ 
\hline
\textbf{S-POP}  & 0.03545 & 0.12288 & 0.15136 & 0.03545 & 0.06877 & 0.07218 & 0.12645 & 0.17678 & 0.18881 & 0.05014 & 0.36861 & 0.41405 \\
\textbf{AR}  & 0.09326 & 0.23779 & 0.29666 & 0.09326 & 0.14436 & 0.15269 & 0.12925 & 0.31183 & 0.40665 & 0.05395 & 0.04773 & 0.05782 \\
\textbf{SR}  & 0.09684 & 0.23781 & 0.38576 & 0.09684 & 0.14695 & 0.16186 & 0.13818 & 0.34088 & 0.55249 & 0.05339 & 0.04686 & 0.04941 \\
\textbf{VSKNN}  & 0.08387 & 0.18405 & 0.28761 & 0.08387 & 0.11959 & 0.12962 & 0.06223 & 0.13235 & 0.18946 & 0.07252 & 0.06734 & 0.05485 \\
\textbf{SMF}  & 0.06524 & 0.25059 & 0.48033 & 0.06524 & 0.12911 & 0.15239 & 0.13977 & 0.29848 & 0.50202 & 0.03742 & 0.03584 & 0.0342 \\
\textbf{Item2Vec}  & 0.04703 & 0.13127 & 0.28418 & 0.04703 & 0.07685 & 0.09154 & 0.14115 & 0.36694 & 0.67071 & 0.01948 & 0.01507 & 0.01152 \\
\hline
\end{tabular}}
\end{table}

\newpage
\begin{figure*}[t]
    \centering
    \includegraphics[width=\textwidth]{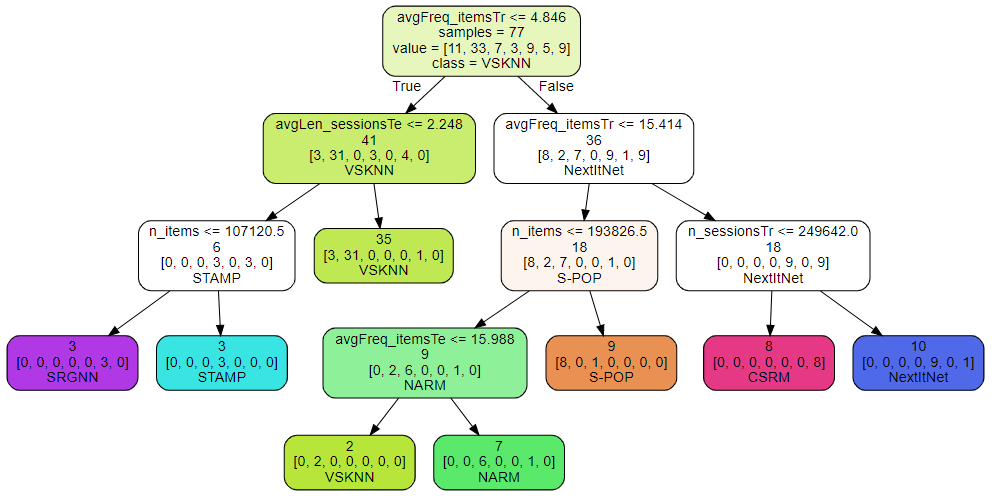}
    \caption{Decision Tree model trained to predict the best model to perform based on dataset characteristics.}
    \label{fig:tree}
\end{figure*}

\clearpage
\subsection{Hyper-parameters' ranges and discretization levels for each model}
\begin{table}
\centering
\caption{List of tuned hyper-parameters ranges and discretization level for each model.}
\label{tab:appendixtab19}
\resizebox{\textwidth}{!}{\begin{tabular}{|c|c|c|c|c|c|c|} 
\hline
\rowcolor[rgb]{0.2,0.2,0.2}  \textbf{\textcolor{white}{Model Name }}  & \multicolumn{6}{c|}{\textbf{\textcolor{white}{Hyper-parameters}} } \\ 
\hline
\multirow{3}{*}{\textbf{S-POP} } & {\cellcolor[rgb]{0.937,0.937,0.937}}\textbf{TopN}  & \multicolumn{1}{c}{} & \multicolumn{1}{c}{} & \multicolumn{1}{c}{} & \multicolumn{1}{c}{} & \multicolumn{1}{c}{} \\ 
\cline{2-2}
 & 10-1000 & \multicolumn{1}{c}{} & \multicolumn{1}{c}{} & \multicolumn{1}{c}{} & \multicolumn{1}{c}{} & \multicolumn{1}{c}{} \\ 
\cline{2-2}
 & Scale: Linear - Step: 10 & \multicolumn{1}{l}{} & \multicolumn{1}{l}{} & \multicolumn{1}{l}{} & \multicolumn{1}{l}{} & \multicolumn{1}{l}{} \\ 
\hhline{|--~~~~~}
\multirow{3}{*}{\textbf{AR} } & {\cellcolor[rgb]{0.937,0.937,0.937}}\textbf{Pruning}  & \multicolumn{1}{c}{} & \multicolumn{1}{c}{} & \multicolumn{1}{c}{} & \multicolumn{1}{c}{} & \multicolumn{1}{c}{} \\ 
\cline{2-2}
 & 0-10 & \multicolumn{1}{c}{} & \multicolumn{1}{c}{} & \multicolumn{1}{c}{} & \multicolumn{1}{c}{} & \multicolumn{1}{c}{} \\ 
\cline{2-2}
 & Scale: Linear - Step: 1 & \multicolumn{1}{l}{} & \multicolumn{1}{l}{} & \multicolumn{1}{l}{} & \multicolumn{1}{l}{} & \multicolumn{1}{l}{} \\ 
\hhline{|---~~~~}
\multirow{3}{*}{\textbf{SR} } & {\cellcolor[rgb]{0.937,0.937,0.937}}\textbf{Pruning}  & {\cellcolor[rgb]{0.937,0.937,0.937}}\textbf{Weighting}  & \multicolumn{1}{c}{} & \multicolumn{1}{c}{} & \multicolumn{1}{c}{} & \multicolumn{1}{c}{} \\
\hline
\end{tabular}}
\end{table}

\end{document}